\documentclass[emulateapj,natbib]{emulateapj}

\begin{document}
\title{Keck/NIRC2 Imaging of the Warped, Asymmetric Debris Disk Around HD 32297}
\author{Thayne Currie\altaffilmark{1},  Timothy J. Rodigas\altaffilmark{2},
John Debes\altaffilmark{3},
Peter Plavchan\altaffilmark{4},
Marc Kuchner\altaffilmark{1},
Hannah Jang-Condell\altaffilmark{5},
David Wilner\altaffilmark{6},
Sean Andrews\altaffilmark{6},
Adam Kraus\altaffilmark{7},
Scott Dahm\altaffilmark{8},
Thomas Robitaille\altaffilmark{9}
}

\altaffiltext{1}{NASA-Goddard Space Flight Center}
\altaffiltext{2}{Steward Observatory, University of Arizona}
\altaffiltext{3}{Space Telescope Science Institute}
\altaffiltext{4}{NEXSCI, California Institute of Technology}
\altaffiltext{5}{University of Wyoming}
\altaffiltext{6}{Harvard-Smithsonian Center for Astrophysics}
\altaffiltext{7}{Institute for Astronomy, University of Hawaii}
\altaffiltext{8}{Keck Observatory}
\altaffiltext{9}{MPIA-Heidelberg}
\begin{abstract}
We present Keck/NIRC2 $K_{s}$ band high-contrast coronagraphic imaging of the luminous debris disk around 
the nearby, young A star HD 32297 resolved at a projected separation of $r$ = 0.3--2.5\arcsec{} ($\approx$ 35--280 AU).  
The disk is highly warped to the north and exhibits a complex, ``wavy" surface brightness profile interior to 
$r$ $\approx$ 110 AU, where the peaks/plateaus in the profiles are shifted between the NE and SW disk lobes.
The SW side of the disk is 50--100\% brighter at 
$r$ = 35--80 AU, and the location of its peak brightness roughly coincides with the disk's mm emission peak.
Spectral energy distribution modeling suggests that HD 32297 has at least two dust populations that may originate 
from two separate belts likely at different locations, possibly at distances coinciding with the surface brightness peaks. 
A disk model for a single dust belt including a phase function with two components
 and a 5--10 AU pericenter offset explains the disk's warped structure and reproduces 
some of the surface brightness profile's shape (e.g. the overall ``wavy" profile, the SB peak/plateau shifts) 
but more poorly reproduces the disk's brightness asymmetry and the profile at wider separations ($r$ $>$ 110 AU).  
Although there may be 
alternate explanations, agreement between the SW 
disk brightness peak and disk's peak mm emission is consistent with
an overdensity of very small, sub-blowout-sized dust and large, 0.1--1 mm-sized grains at $\approx$ 45 AU 
tracing the same parent population of planetesimals.  New near-IR and 
submm observations may be able to clarify whether even more complex grain scattering properties or 
dynamical sculpting by an unseen planet are required to explain HD 32297's disk structure.
\end{abstract}
\section{Introduction}
Debris disks are signposts of planets and planet formation
\citep[e.g.][]{Wyatt2008,KenyonBromley2008}.  
Supporting this picture, the two stars with independently confirmed, directly imaged planetary systems 
 HR 8799 and $\beta$ Pictoris \citep{Marois2008,Marois2010,Currie2011a,Lagrange2010} 
are surrounded by luminous debris disks \citep{Smith1984,Rhee2007,Su2009}.
Similarly, Fomalhaut has a candidate planetary companion located just interior to the 
star's bright debris ring \citep{Kalas2008}. 

In the absence of a directly imaged planet, resolved imaging of debris disks may provide indirect evidence for 
a massive planet's existence, may help constrain the unseen planet's properties, and 
thus can help identify promising targets for future direct imaging \citep[e.g.][]{Wyatt1999}.  
For example, the inclined or ``warped" component of $\beta$ Pictoris's debris disk \citep{Heap2000,Golimowski2006} 
is likely due to the directly imaged planet \citep{Augereau2001,Dawson2011} and also provides an estimate for the planet's mass 
independent of planet cooling models \citep{Lagrange2010}.  Dynamical sculpting by a planet/planets 
may explain the sharp inner edge and pericenter offset of Fomalhaut's debris ring \citep{Kalas2005a,Quillen2006,Kalas2008}.
Other debris disk structures may be due to non-planet processes, in particular interactions 
with the interstellar medium or perturbations from a nearby star, as has been proposed to 
explain images of disks around HD 15115 and HD 61005 \citep[e.g.][]{Kalas2007,Hines2007}.

The nearby \citep[$d = 112^{+12}_{-10}$ $pc$;][]{vanLeeuwen2007} A5 star HD 32297 is another example of a 
young star surrounded by a luminous, spatially-resolved, debris disk.
At 30 Myr old \citep{Kalas2005a}, it is roughly coeval with HR 8799 and may probe 
debris disk evolution at a stage just after they are 
 most collisionally active \citep{KenyonBromley2008,Currie2008,Currie2009}.
Like $\beta$ Pic and HR 8799, HD 32297 has a large infrared (IR) excess emission due to circumstellar 
dust first identified from \textit{IRAS} data.  \citet{Schneider2005} thus
selected HD 32297 for \textit{Hubble Space Telescope}(\textit{HST}) NICMOS (F110W) coronagraphic imaging and 
resolved the disk out to an angular distance (from the star) of $\sim$ 3.3\arcsec{} ($\sim$ 400 AU).
 HD 32297 was subsequently resolved in the optical \citep{Kalas2005b}, near-IR \citep[1.6--2.2 $\mu m$,][]{Debes2009,Mawet2009}, 
thermal infrared \citep[10--20 $\mu m$,][]{Moerchen2007,Fitzgerald2007}, and millimeter \citep[1.3mm][]{Maness2008}.

Previous work has claimed that HD 32297's disk structure is shaped by planet sculpting as well as non-planet processes.
\citet{Debes2009} identified an asymmetry in the disk scattering efficiency between the northeast and southwest sides 
\citep[see also][]{Kalas2005b}.  
They argued that ISM sculpting explains this feature much like it explains some properties of the HD 15115 and HD 61005 disks.  
\citet{Schneider2005} identified a brightness asymmetry between the two disk sides,
 a feature consistent with sculpting by a massive planet \citep[see also][]{Maness2008}.  

  The two mechanisms, ISM sculpting and planets, are not mutually exclusive.  New images of the HD 15115 and HD 61005 disks reveal 
cleared inner regions and/or pericenter offsets, both of which are plausibly due to a planetary companion \citep{Rodigas2012, Buenzli2010}.  
Additionally, multiple debris belts, scaled analogues to the solar system's asteroid belt and Kuiper belt, are also likely planet 
signposts and may reside around HD 61005, HD 15115, and HD 32297 \citep{Fitzgerald2010, Maness2008,Rodigas2012}.  

To determine which mechanisms are responsible for shaping HD 32297's debris disk structure, we need new, high signal-to-noise 
images with which to derive precise disk properties.  Although \citet{Schneider2005} identify a disk brightness asymmetry consistent 
with planet sculpting, they caution that the disk brightness measurements close to the coronagraphic spot (r $\sim$ 0.3--0.4\arcsec{}) 
which provide the basis for this asymmetry are highly uncertain.  PSF subtraction errors due to the completely opaque NICMOS 
coronagraphic spot may limit our ability to conclusively identify disk structure at these small, speckle-dominated separations.
Moreover, if the asymmetry identified a planet-induced density structure, it should align with the mm emission peak 
\citep{Maness2008}.  However, 
it is not clear whether these asymmetries are aligned and thus whether they identify small and large grains originating 
from the same parent population.
The Palomar/$K_{s}$ image from \citet{Mawet2009} has limited spatial resolution.  
While they did recover \citeauthor{Schneider2005}'s brightness asymmetry, 
higher spatial resolution observations could confirm and help clarify the physical 
origin of this and other features.  For example, new data could 
 identify 
breaks in the disk brightness profile that may reveal evidence for the multiple debris belts 
inferred from modeling unresolved IR data. 

To further clarify the nature of the HD 32297 debris disk, we present new coronagraphic imaging obtained 
at $K_{s}$ ($\sim$ 2 $\mu m$) with the Keck telescope on Mauna Kea, resolving the disk at angular separations of 
 0.3--2.5\arcsec{}.  \S 2 describes our observations and extraction of the disk images.  
In \S 3, we investigate basic disk properties (position angle, full-width half-maximum, and surface brightness) as a function of 
angular separation from the star.  We then combine imaging with new, unresolved broadband photometry from the 
\textit{Spitzer Space Telescope} and the \textit{WISE} satellite to constrain the disk structure and identify the location(s) of the 
disk emission (\S 4).  Finally, we compare our analyses to those from previous work 
on HD 32297 (\S 5) and investigate the physical mechanisms responsible for sculpting the disk emission (\S 6).  

\section{NIRC2 Data}
\subsection{Observations}
We imaged HD 32297 on UT November 20, 2011 with the NIRC2 camera mounted 
on Keck II using the $K_{s}$ filter ($\lambda_{o}$ = 2.16 $\mu m$) in 
the narrow camera mode \citep[9.952 mas/pixel][]{Yelda2010} with correlated 
double sampling.  The Keck AO system delivered diffraction limited images 
with a FWHM of $\sim$ 4.9 pixels ($\sim$ 49 mas).
To enhance our ability to extract the HD 32297 disk emission from the bright 
stellar halo, we centered the star behind the 0.6\arcsec{} diameter, partially-transmissive 
coronagraphic spot and used the "large hex" pupil plane mask.  

Our HD 32297 data consist of coadded 30s exposures with a cumulative integration time of 
20 minutes and were taken through transit (HA = [-0.22,0.35]) in ``vertical angle" or 
\textit{angular differential imaging} mode \citep[ADI,][]{Marois2006}.   Over the course 
of these exposures, the parallactic angle changed by 36 degrees.  
While light cirrus caused fluctuations in the source transmission (as measured 
from the PSF core behind the coronagraphic mask) on the order of $\sim$ 10\%, observing 
conditions were otherwise stable. 

To flux calibrate the data, we measure the flux of the PSF core of HD 32297 
as viewed through the partially transmissive coronagraphic mask and corrected 
for the extinction through the mask.  To determine the extinction, we use 
Keck/NIRC2 $K_{s}$ observations of HD 15115 taken in August 2011 in photometric conditions 
with the same coronagraphic spot size that were flux-calibrated with a standard star HD 3029.  
From the August data, we measure the extinction through the 0.6\arcsec{} 
spot to be 6.91 $\pm$ 0.15 mags, slightly less than but comparable to 
extinction estimates in $K_{s}$ for 1--2\arcsec{} spot diameters \citep{Metchev2009}.  
Uncertainties in the aperture correction inside the coronagraphic spot due to scattered light 
from the spot edge limit the precision of our estimate.  We did observe a photometric standard (p486r) 90 minutes 
after our HD 32297 imaging sequence.  Although conditions became highly 
variable during the standard star observations due to patchy cloud coverage, our absolute 
flux calibration implied from the least extincted standard star frames 
agrees with that derived from the coronagraph transmission to 
within 0.3 magnitudes.  

\subsection{Image Processing}
Basic image processing and high-contrast imaging techniques follow
methods outlined in \citet{Currie2010,Currie2011a,Currie2011b}.
We employ standard dark subtraction and flatfielding corrections, identify and 
interpolate over hot/cold pixels, and apply the distortion correction determined 
from \citet{Yelda2010}.  For image registration, we exploit the fact that 
the PSF core is visible through the coronagraphic mask and is unsaturated.  
We register each image to subpixel ($\sigma_\mathrm{cen}$ $\approx$ 0.1 pixels) accuracy by 
determining the centroid position of the first image in the sequence 
and then determining the relative offset between the 
reference image and subsequent images by solving for the peak in the cross-correlation 
function for each image pair.  Finally, we subtract off a radial profile to remove the smooth seeing halo and 
make a second pass through the images to identify remaining bad pixels.

To extract a detection of the HD 32297 disk,
we adopt the \textit{Locally Optimized Combination of Images}
approach \citep[LOCI][]{Lafreniere2007}, using an updated version of the LOCI-based code employed in 
\citet{Currie2011b}, which will be detailed later \citep[][T. Currie, 2012 in prep.]{Currie2012}.  
Following \citet{Thalmann2011}, we reduce the data using ``conservative" LOCI 
settings more appropriate for extended sources (i.e. disks, not planets), 
where we consider 
rotation gaps $\delta$ of 1.5--5$\times$ FWHM, optimization areas ($N_{A}$) of 
1000--3000$\times$ the FWHM area, optimization geometries $g$ of 0.5--2, and subtraction annuli ($dr$) of 
5--10 pixels wide \citep[see ][for more details]{Lafreniere2007}.  
To determine the signal-to-noise per resolution element of our disk detection, we convolve the image 
with a beam size equal to $\sim$ 1 FWHM and compare pixel counts to the standard 
deviation of counts within a 1 FWHM-wide annulus\footnote{The focus of this paper is imaging 
and characterizing the HD 32297 debris disk.  While we do not present data reductions with 
methods more optimized for point source detection, we plan to do so in a future work}.  
Finally, we correct for photometric/astrometric biases inherent in LOCI-based processing 
by imputing fake disks into each registered image and comparing the predicted and measured
 disk properties (e.g. surface brightness, full-width half maximum, position angle).  Our 
method follows that first developed by \citet{Rodigas2012} and is described in full in 
the Appendix.

\subsection{HD 32297 Disk Image}
Figure \ref{images1} shows the image and signal-to-noise map for 
our ``conservative" LOCI reduction assuming a rotation gap criterion of $\delta$ $\ge$ 2.5, 
$N_{A}$ = 3000, $g$ = 1, and $dr$ = 10, which 
balances our ability to detect the disk at small separations by attenuating speckles (favoring 
smaller $\delta$, $N_{A}$) but without oversubtracting the disk (favoring larger $\delta$, $N_{A}$).  
As evidenced by the signal-to-noise map, we detect the disk at SNR $\ge$ 3 from r = 2.5\arcsec{} 
all the way to the edge of the coronagraphic spot at 0.3\arcsec{}.  The peak signal-to-noise per pixel 
is $\sim$ 19.  The disk emission along the midplane is more than 10$\sigma$ significant 
at 0.85--1.4\arcsec{}, while some regions of disk emission at 0.3--0.5\arcsec{} on the SW side are still more than 
5-$\sigma$ significant.
Visually inspecting the image and signal-to-noise map reveals significant disk structure.
Most notably, the disk emission traces a distinct ``bow" shape, where the 
disk position angle clearly changes with radial separation.

Furthermore, the disk exhibits a significant brightness asymmetry at small separations ($r$ $<$ 0.7\arcsec{}).
Figure \ref{mawet} redisplays our disk image with a different color stretch to better illustrate the 
brightness differences between the two sides of the disk.  Interior to $r$ = 0.35\arcsec{} (identified with a circle), the SW side is 
significantly brighter than both the NE side as well as any pixel value for residual speckles located at different azimuthal 
separations; the NE side has a peak brightness only slightly larger than the brightest speckle.  Exterior to this separation, 
there are no residual speckles as bright as either side of the disk, and the SW is still clearly brighter at least out to 
$r$ $\approx$ 0.6\arcsec{} (yellow/red region on the SW side). 

Our Keck $K_{s}$ image agrees well with the previous best $K_{s}$ results, which were
 obtained with the extreme-AO \textit{Well-Corrected Sub-Aperture} 
on Palomar presented by \citet{Mawet2009} using classical (not ADI) imaging.  Convolved to the Palomar/WCS 
beamsize (Figure \ref{mawet}, bottom panel), our image strongly resembles that of \citeauthor{Mawet2009}'s.  
The disk appears highly asymmetric with the SW side being 
brighter extending all the way to the coronagraphic spot 
(0.3\arcsec{} in our image compared to 0.4\arcsec{} in theirs).  As with the \citet{Mawet2009} data, 
the NE side appears fainter and truncated.  Moreover, the brightest portion of the disk on the SW side 
roughly overlaps with the mm continuum peak \citep{Maness2008}.

\section{Analysis}
Here, we investigate the HD 32297 disk geometry and 
surface brightness profile, following methods similar to those 
described in \citet{Rodigas2012}.  We perform analysis on the 
image shown in Figure \ref{images1}.  
The disk properties we report are corrected for photometric/astrometric 
biases inherent in LOCI processing (see Appendix).
\subsection{Disk FWHM}
To better assess the HD 32297 disk morphology, we 
measure the disk FWHM perpendicular to the disk's 
major axis as a function of stellocentric distance.
First, we identify the brightest pixels at each radial separation for the NE and SW lobes, respectively.  
Next, we place a 5 pixel by 21 pixel box centered on the brightest pixel and 
sum up the counts/pixel along each row of the box, producing a 1D array of 21 values.  
Finally, we fit a Gaussian to this array, which yields the disk midplane location and 
the disk FWHM at that location.   

Figure \ref{diskpafwhm} displays the disk FWHM as a function of stellocentric distance for the NE (purple) 
and SW (green) sides.  The errors correspond to the residuals of the Gaussian fits divided by the 
``throughput" for the disk FWHM as determined in the Appendix.
On both sides, the disk FWHM steadily decreases from $\sim$ 0.25\arcsec{} at r = 1.5\arcsec{}
 to $\approx$ 0.15\arcsec{} at r = 1\arcsec{}.  Interior to $r$ = 1\arcsec{}, the disk FWHM
 fluctuates about a constant value, though this behavior is likely due to the difficulty of 
fitting a Gaussian profile in speckle-dominated regions, not bona fide clumpy structure.
  Beyond $r$ = 1.5\arcsec{}, where 
the disk emission approaches the photon noise limit, the FWHM estimates fluctuate 
wildly.

\subsection{Disk Position Angle}
\begin{figure}
\centering
\includegraphics[scale=0.6,trim=31mm 0mm 35mm 0mm,clip]{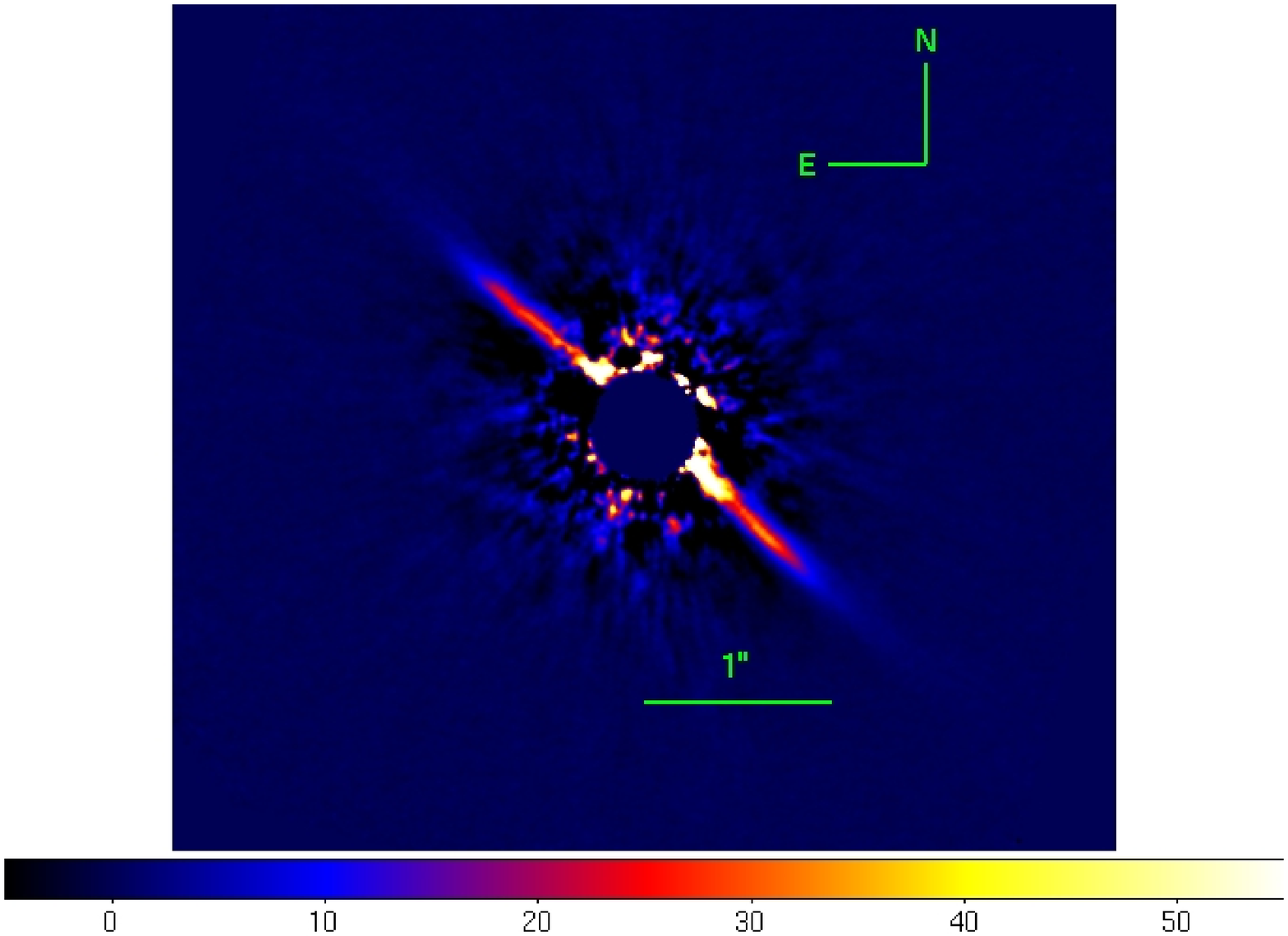} 
\\
\includegraphics[scale=0.6,trim=31mm 0mm 35mm 0mm,clip]{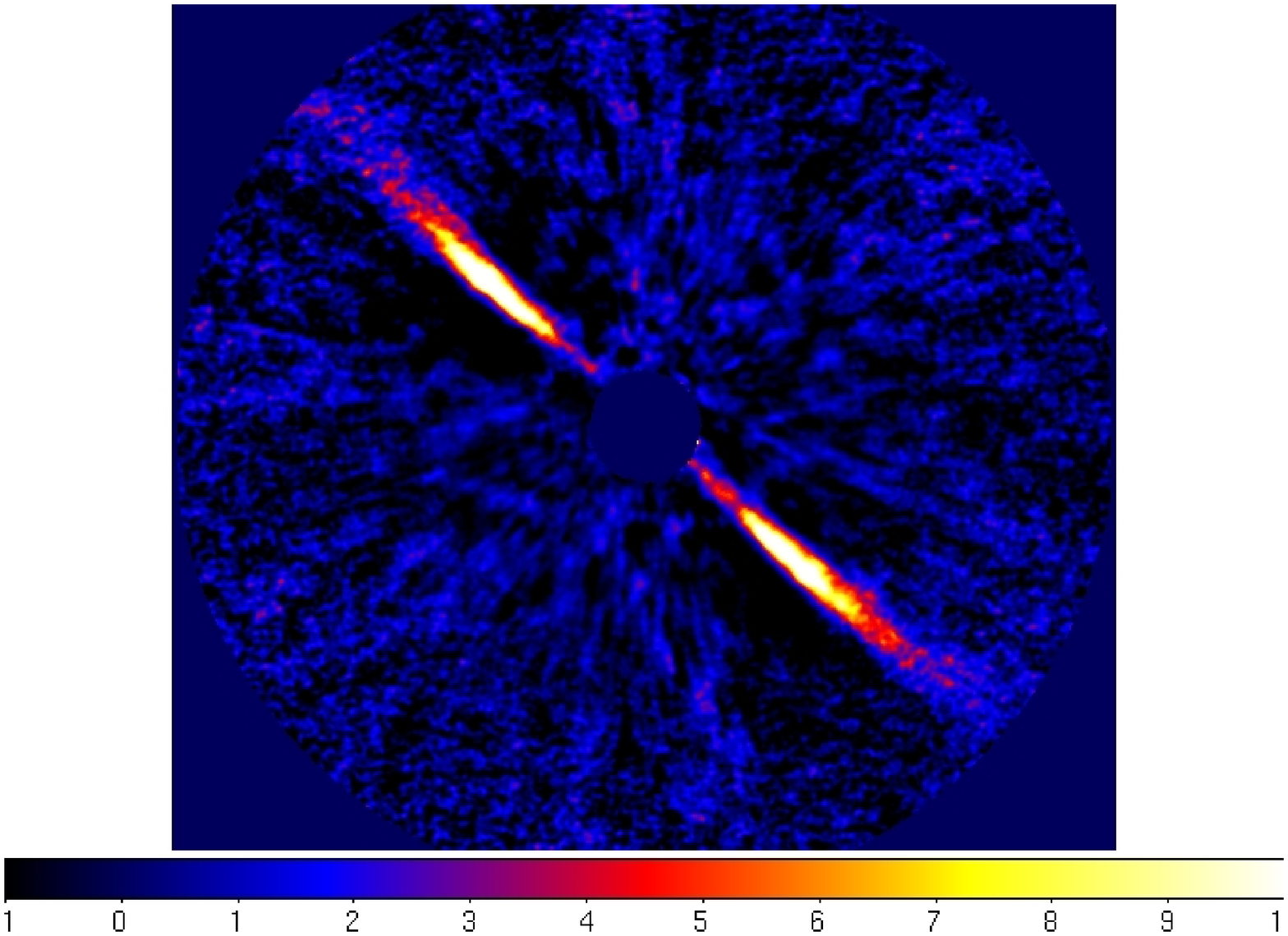}
\caption{Reduced image (top) and signal-to-noise map (bottom) for our NIRC2 HD 32297 data.  The colorbar 
depicting units for the image are in counts, whereas they range from 0 to 9$\sigma$ for the signal-to-noise 
map.  The central dark region identifies the coronagraphic spot ($r$ = 0.3\arcsec{}).  The panels have the same size scale.}
\label{images1}
\end{figure}

\begin{figure}
\centering
\vspace{-0.1in}
\includegraphics[scale=0.6,trim=31mm 0mm 35mm 0mm,clip]{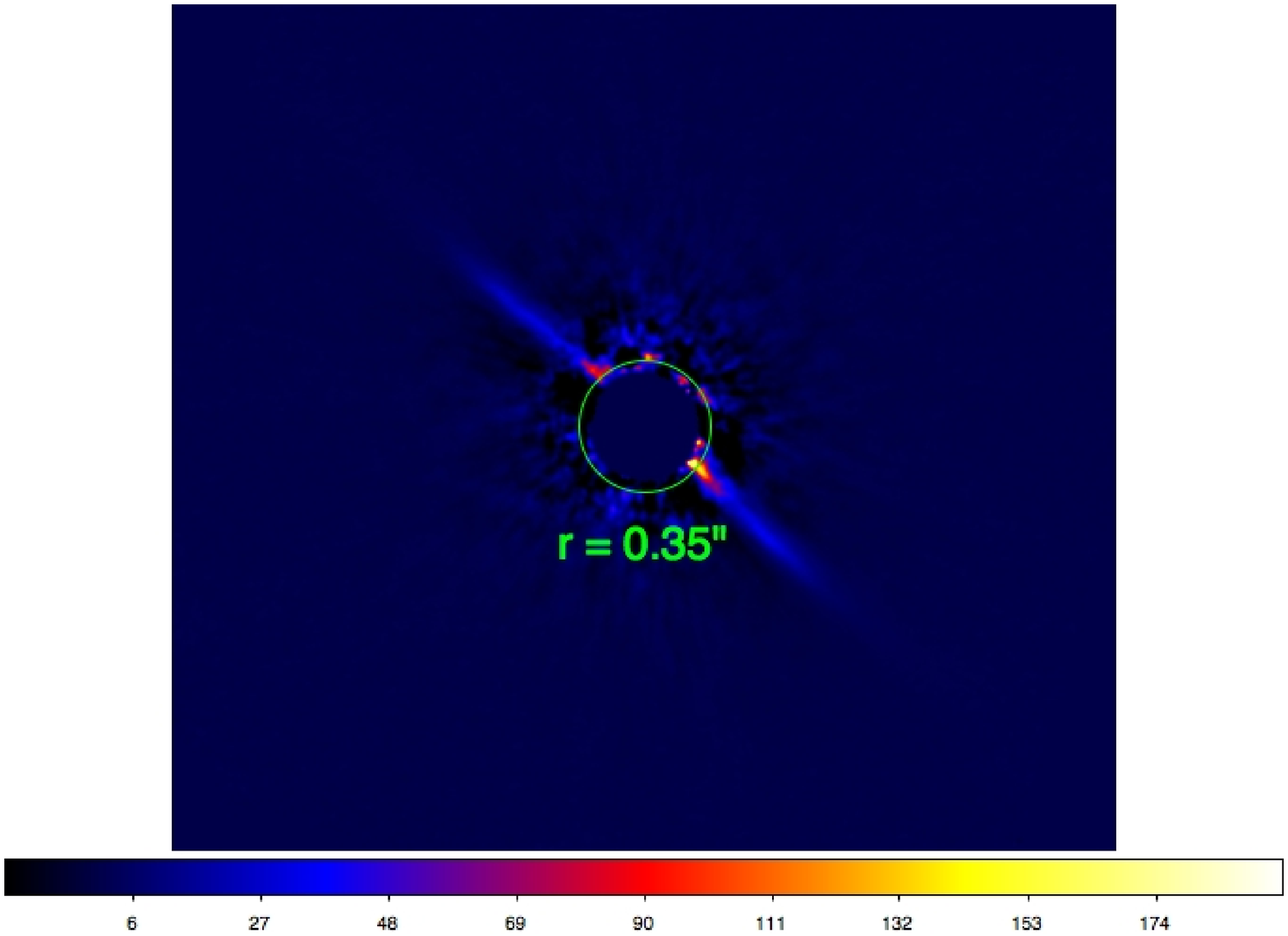}
\vspace{-0.1in}
\includegraphics[scale=0.6,trim=31mm 0mm 35mm 0mm,clip]{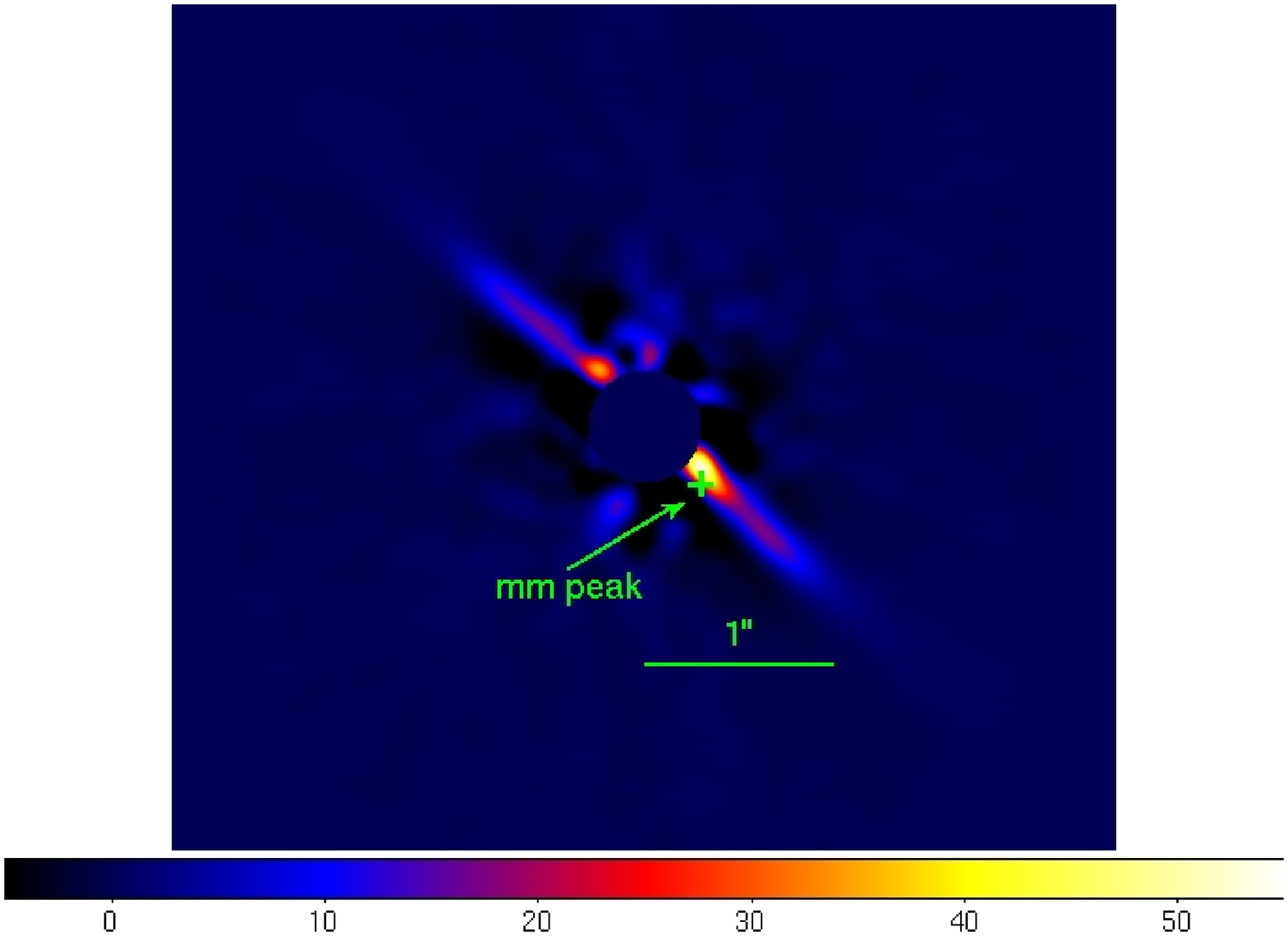}
\vspace{-0.1in}
\caption{(Top) Our image from the top panel of Figure 1 redisplayed with a different color stretch to 
better illustrate the 
significance of our disk detection at the smallest separations ($r$ =0.3--0.6\arcsec{}) and the brightness asymmetry.
(Bottom) Our image resampled to the same spatial resolution as the Palomar/WCS image from \citet{Mawet2009}.  
The green cross identifies the position (and positional uncertainties) of the peak brightness in the millimeter 
\citep{Maness2008}.  Both panels are displayed in units of counts.} 
\label{mawet}
\end{figure}

\begin{figure}
\centering
\includegraphics[scale=0.5,trim=0mm 0mm 0mm 0mm,clip]{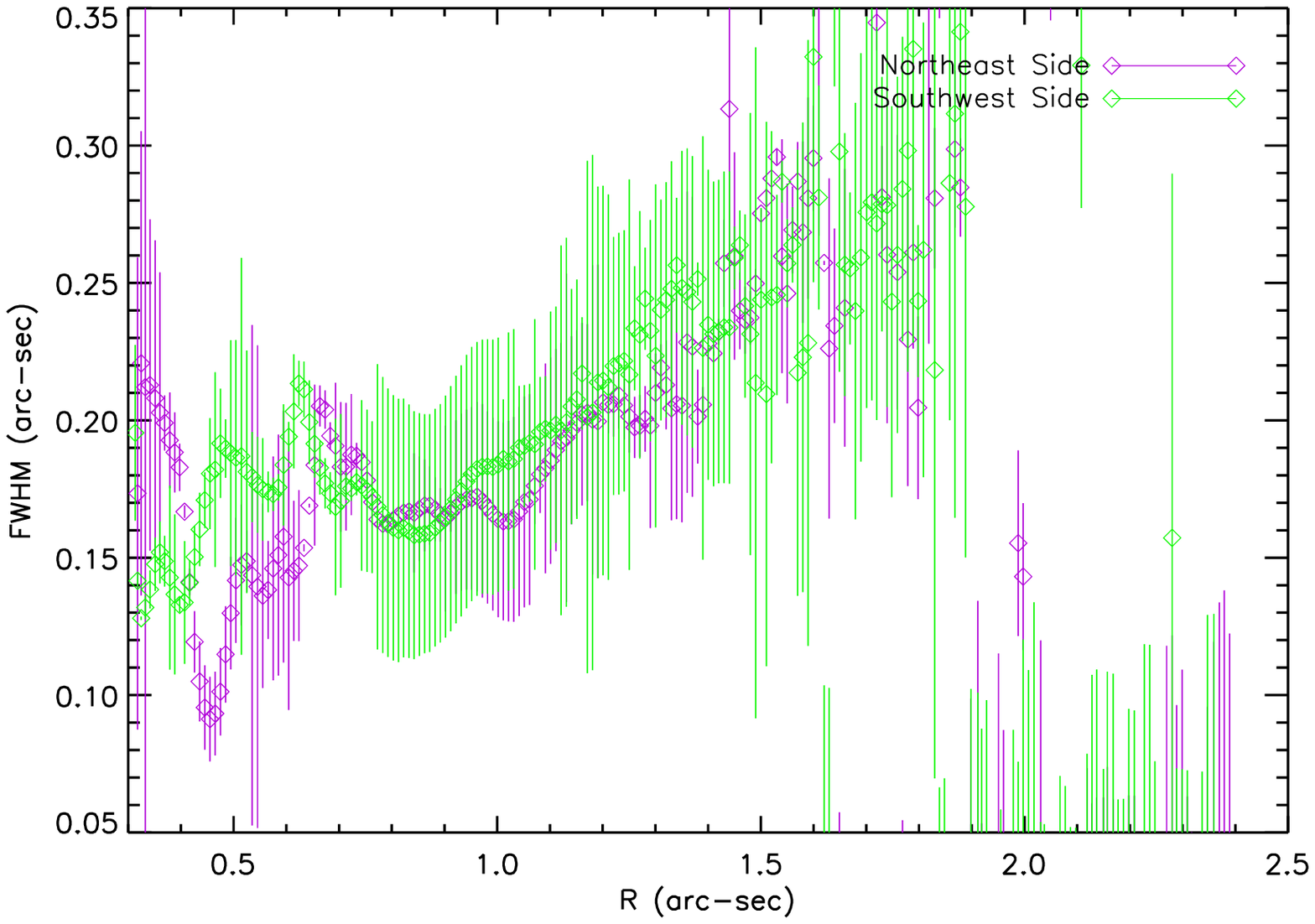}
\includegraphics[scale=0.5,trim=0mm 0mm 0mm 0mm,clip]{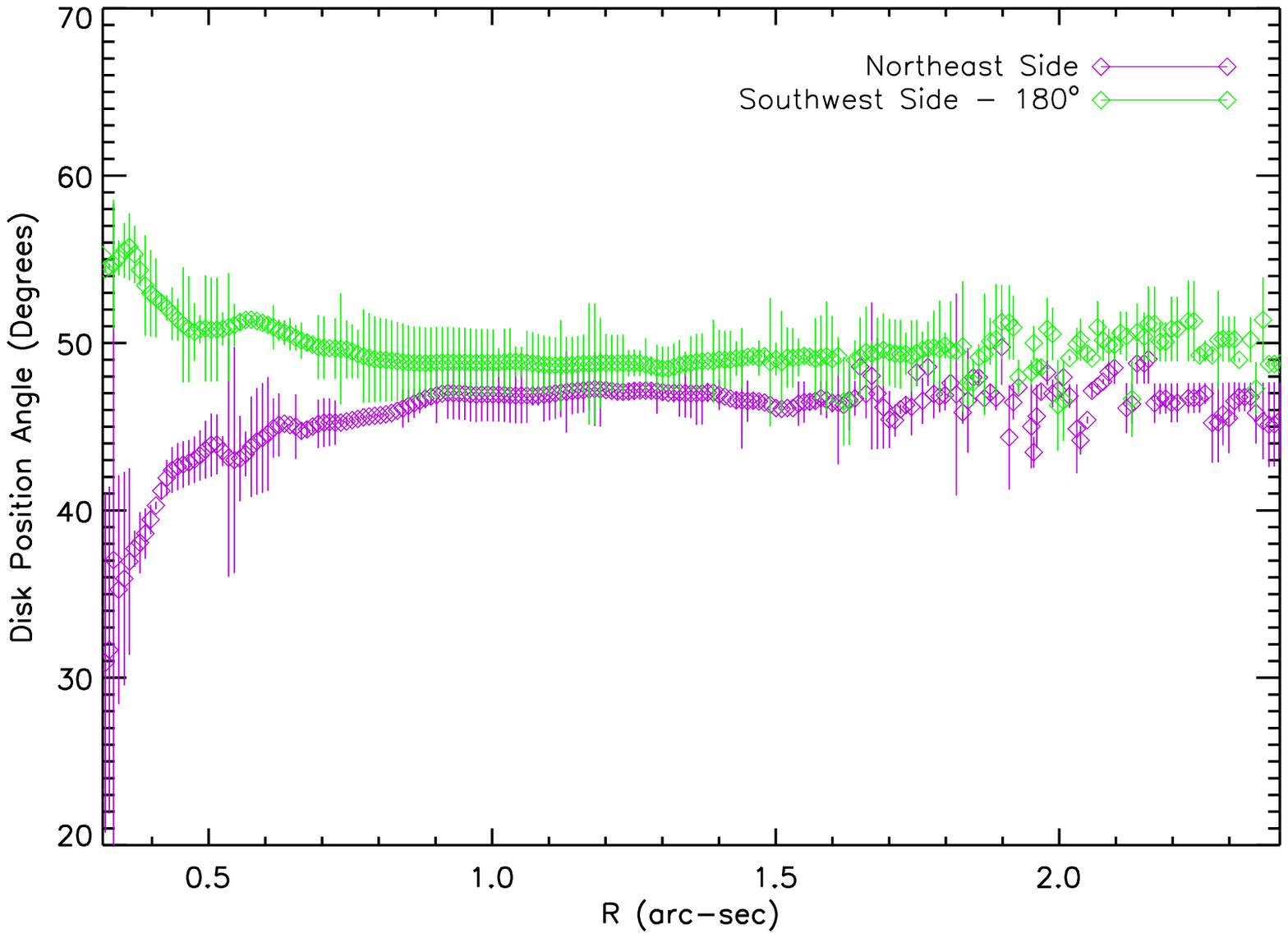}
\caption{Disk FWHM (left) and position angle vs. angular separation for the NE and SW sides of the disk.  The disk 
narrows at smaller angular separations.  The two sides of the disk are offset in position angle by $\sim$ 3--4$^{\circ}$; 
the disk curves towards the north starting at $r$ = 0.9\arcsec{}.}
\label{diskpafwhm}
\end{figure}

To quantify the ``bow" structure easily seen in Figure \ref{images1} 
and identify any additional warping, we 
calculate the disk position angle as a function of stellocentric distance for both 
the NE and SW sides using the disk midplane pixel locations from the Gaussian fits described 
in \S 3.1. 
The position angle uncertainty at each 
radius results from the difference between the Gaussian-fitted disk midplane location at the 
radius and a midplane location defined by the brightest pixel.
Here we formally assume a systematic uncertainty of 0.009$^{\circ}$ from the 
\citeauthor{Yelda2010} astrometric calibration, although fitting errors always dominate.  

Beyond r = 0.9\arcsec{}, 
both sides of the disk maintain a constant position angle, although they are misaligned 
by 3--4 degrees.  Between 0.3\arcsec{} and 0.9\arcsec{}, though, the disk 
emission on both sides curves towards the north, with the position angle decreasing on the NE side by more than 20$^{\circ}$   
and increasing by $\sim$ 10$^{\circ}$ on the SW side.  On the SW side, this curvature 
is not continuous, leveling off at 232$^{\circ}$ at 0.5--0.6\arcsec{} before resuming 
at smaller separations.  Beyond $r$ = 1.6\arcsec{}, the disk exhibits no obvious curvature/warping, though 
photon noise degrades the precision of our estimates.
 
\subsection{Disk Surface Brightness Profile}
To calculate the disk surface brightness (SB), we 
follow \citet{Rodigas2012} and
determine the median surface brightness in $mJy$/$arcsec$$^{2}$ in a 
18-pixel diameter circular aperture.
  We determine uncertainties in the median disk surface brightness (also in $mJy$/$arcsec$$^{2}$) 
at a given angular separation in a way analogous to that which we used to determine the 
disk SNR/pixel.  
Specifically, at each pixel radius corresponding to a Gaussian-fitted disk 
midplane location we calculate the uncertainty in SB within 
non-overlapping circular apertures identical in size to that which we use to 
determine the disk SB and covering all azimuthal angles exterior 
to the disk.  We adopt the standard deviation of these SB estimates as the 
uncertainty in the disk SB at each radius.

\begin{figure}
\centering
\includegraphics[scale=0.7,trim=0mm 0mm 0mm 0mm,clip]{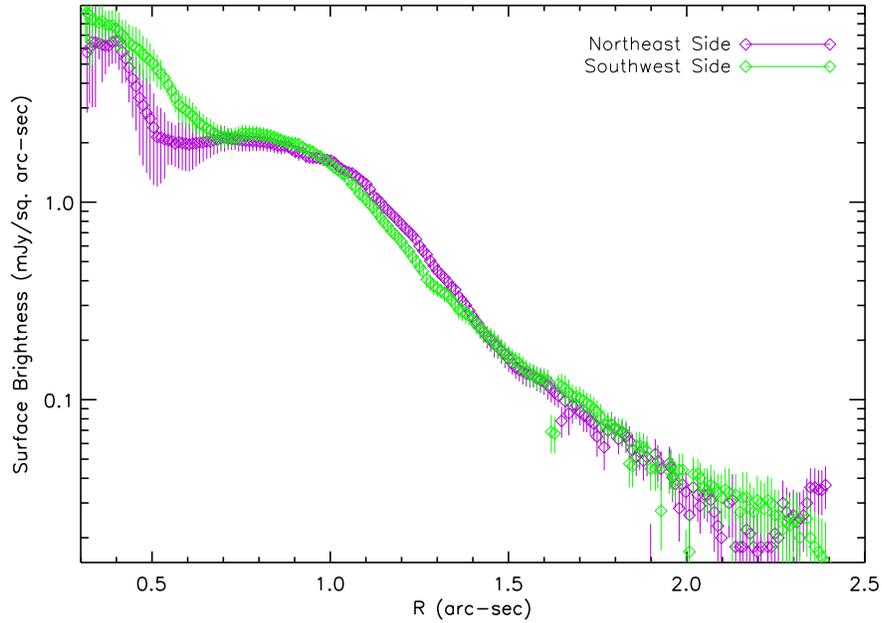}
\caption{Surface brightness profiles for the two sides of the disk.  In agreement with previous work \citep[e.g.][]{Schneider2005}, the 
disk exhibits power law breaks at $r$ = 1.5\arcsec{} and $r$ = 1.1\arcsec{}.  We identify a strong jump in surface brightness 
starting at $r$ $\approx$ 0.5--0.7\arcsec{}.  }
\label{disksb}
\end{figure}

Figure \ref{disksb} displays the surface brightness profiles for both sides of the disk.  
On both sides, the disk steadily brightens from $r$ = 2.5\arcsec{} to $r$ = 1\arcsec{}.  
However, at smaller separations the profiles change dramatically.  On the NE side, 
the disk has a roughly constant brightness at 0.5--0.9\arcsec{}
 before brightening from 3 to 6 $mJy$/$arcsec^{2}$ at 0.5\arcsec{} to 0.4\arcsec{}.  The SW side of the disk displays a similar behavior: 
a nearly constant brightness at 0.7--0.9\arcsec{}, a possible slight dip in brightness at 0.7\arcsec{}, 
and a sharp jump in brightness by a factor of 4 from 0.7\arcsec{} to 0.4\arcsec{}.  
Thus, the disk SB profile appears ``wavy" interior to $r$ = 1\arcsec{}.

The two sides of the disk differ slightly in some other important respects.  
The locations of the peaks/plateaus closer to the star for the NE side 
than for the SW side.  The shifted profiles are consistent with the dust ring being 
located at different stellocentric distances between the NE and SW sides: a pericenter offset.  
The NE side also plateaus and may drop slightly at $r$ $\sim$ 0.3--0.35\arcsec{}.
Conversely, on the SW side the disk continues to brighten all the 
way to our inner working angle of 0.3\arcsec{}.  

Moreover, our analysis confirms evidence for a brightness asymmetry between the 
NE and SW sides.
These differences are significant even at small separations where we detect the 
disk at a lower SNR.  Interior to 0.6\arcsec{}, the median uncertainty in the disk 
SB is $\sim$ 2.8 on the NE side and 4.6 on the SW side.  While these uncertainties are large, 
they are significantly smaller than the brightness differences at $r$ = 0.5--0.6\arcsec{}, where 
the SW side is 2--3 times brighter.  At these angular separations, the +/- 1-$\sigma$ 
range in SB between the two sides do not overlap.  The SW side of the disk is 50\% brighter 
at r $\sim$ 0.3--0.4\arcsec{}, although the large error bars for the NE side SB make this 
difference less statistically significant.

We model the surface brightness profiles over $r$ = 0.3--2.5\arcsec{}
to derive power law indices $a$ and $b$ assuming a functional form of 
$f(X)$ = $a$X$^{b}$.  
As a first guess, we take the logarithm of the surface brightness, adopt uniform 
weighting to each datapoint, and estimate the power indices by fitting a straight 
line and deriving the slope.   We then refine our power law index estimates 
by performing a Levenberg-Marquardt minimization, while considering measurement 
errors.  To more precisely estimate 
the locations of the power law breaks, we perform the above steps iteratively, varying
the break locations and adopting ones that minimize the reduced $\chi^{2}$
between the data and the model.  

Table \ref{plawtable} summarizes our results.  We fail to identify any power law to either 
lobe that quantitatively fits the entire radial extent of the disk.  For the NE (SW) side, 
the goodness-of-fit value ($R^{2}$) declines 
 to zero at $r_{in}$ $\sim$ 1.0\arcsec{} (0.9\arcsec{}).  
Assuming a single power law for both sides, 
we derive best-fit coefficients of $a$ = 2.13, $b$ = -6.01 at $r$ = 1.05--2.5\arcsec{} 
for the NE side and $a$ = 1.72, $b$ = -5.49 for the SW side at $r$ = 0.95--2.5\arcsec{}.  
However, the goodness-of-fit criterion for the NE side is low ($R^{2}$ $\sim$ 0.75), indicating 
that at least this side may be best modeled as a broken power law \citep[see also][]{Schneider2005,Debes2009}.

For the NE side, the best-fit indices assuming a broken power law are 
$a$ = 2.23, $b$ = -6.19 at $r$ = 1.1--1.4\arcsec{} and $a$ = 1.31, $b$ = -5.13 at $r$ = 1.4--2.5\arcsec{}.  
For the SW side, the best-fit indices are similar: $a$ = 1.78, $b$ = -5.71 at $r$ = 1--1.6\arcsec{} and 
$a$ = 1.66, $b$ = -5.33 at $r$ = 1.65--2.5\arcsec{}\footnote{While the inner and outer fitted radii reported 
here are the ones for which a power law fit is most applicable, we obtain similar results for slightly 
different radii: e.g. for $r_{in}$ = 0.95\arcsec{} and $r_{out}$ =1.65\arcsec{}.}.  Formally, our values 
for $b$ have a large uncertainty since the goodness-of-fit criterion remains above 0.95 for 
$a$ $\pm$ 20\% and $b$ $\pm$ 0.5.

\begin{figure}
\centering
\plotone{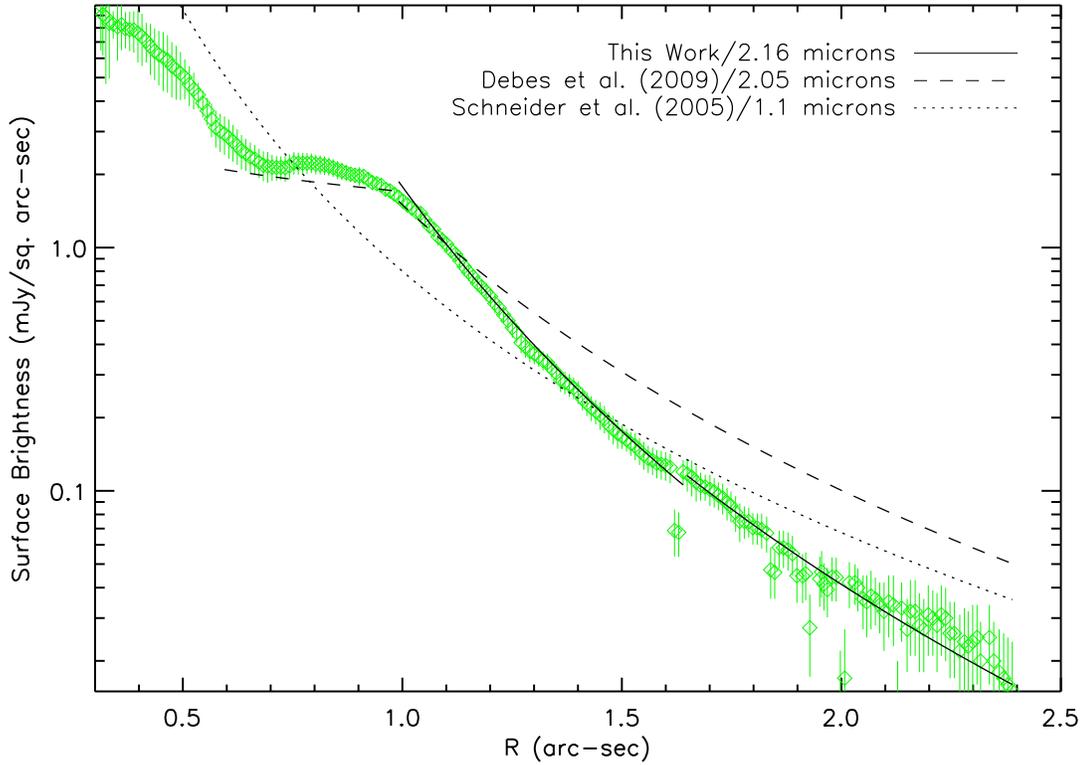}
\caption{Comparisons between our best-fit power laws to the HD 32297 $K_{s}$ surface brightness profile (SW side) 
and fitted power laws for 1.1--2.05 $\mu m$ HD 32297 data from \citet{Schneider2005} and \citet{Debes2009}.  Our fits 
are generally much steeper.  We find that the surface brightness profile interior to $r$ $\sim$ 1\arcsec{} cannot be 
fit by a power law.}
\label{sb_compmodel}
\end{figure}

We can compare our SB profiles to those from \citet{Schneider2005} and 
\citet{Debes2009} to understand how the profiles change with wavelength.
In general, the shape of the SB profiles show good agreement with 
those derived from 1.1 $\mu m$ data by \citet{Schneider2005}.
  Although they do not draw attention to any 
steep increase in SB at $r$ = 0.5--0.7\arcsec{}, their Figure 2 provides 
some evidence for this feature, at least on the SW side.
While we do not find evidence for a steep drop in SB at $r$ = 0.3--0.5\arcsec{} 
for the NE side as they do, the SB does plateau at $r$ = 0.4\arcsec{} and drop slightly.
The 2.05 $\mu m$ profile from \citet{Debes2009} does not extend to 
$r$ $<$ 0.5--0.6\arcsec{}, so we cannot know whether they too find 
evidence for a jump in SB at small separations.
Our profiles are significantly steeper than those from \citet{Schneider2005} and \citet{Debes2009} 
at $r$ $>$ 1 \arcsec{} obtained at shorter wavelengths (Figure \ref{sb_compmodel}).


\section{Debris Disk Modeling}
\subsection{Scattered Light Modeling}
To understand the disk's grain properties and morphology, we compare our 
image to synethetic resolved disk models with 
a range of scattering properties.  
%
We modeled a number density
distribution of dust for the disk in the following cylindrical form, which allows for a variety of disk morphologies:

\begin{eqnarray}
N(r,z) & = & \exp{\left(\frac{r-r_{\rm o}}{2 \sigma_r^2}\right)}\exp{\left(\frac{z}{2 \sigma_z^2}\right)}, r < r_{\rm break} \\
& = & \left(\frac{r}{r_{\rm break}}\right)^{-\beta}, r \ge r_{\rm break}
\end{eqnarray}
where $r_{\rm o}$ can be interpreted as the location of a birth ring of planetesimals
generating a collisional cascade of dust with a characteristic width $\sigma_r$.  We
assume a Gaussian scale height to the disk at all radii, and we allow for a power-law
decrease in dust density beyond some radius $r_{\rm break}$, thus allowing for combinations
of ring-like and disk-like structures.  We place dust particles at various distances
from the star following the density distribution above and distribute them uniformly in 
the azimuthal angle ($\theta$).
To derive the integrated surface brightness, we project the density distribution on a
two dimensional plane after transforming the coordinates of dust particles to account for
inclination and position angle and scale the result to the observed surface brightness of
the disk in the Keck images.  The surface brightness, $F$, of each particle is determined
from its distance from the star and its scattering angle $\omega$ assuming a two component Henyey-Greenstein phase function:

\begin{eqnarray}
F \propto \left(\frac{1}{r^2+z^2}\right)\left[a_1 \frac{1-g_1^2}{(1+g_1^2-2 g_1 \cos{\omega})^{1.5}}+ 
a_2 \frac{1-g_2^2}{(1+g_2^2-2 g_2 \cos{\omega})^{1.5}}\right].
\end{eqnarray}

A multi-component phase function for circumstellar dust may be favored.  For example, the phase
function of zodiacal dust in the Solar System has been modeled with multiple Henyey-Greenstein components,
including a significant backscattering component \citep{hong85}.  Recently, observations of HR~4796A's 
disk surface brightness as a function of scattering angle showed an incredibly flat inferred phase function for its
dust at scattering angles $> 50\arcdeg$ \citep{dalleore11}, and observations of the protoplanetary disk
HD~100546 at multiple wavelengths require forward scattering grains where the phase function becomes
flatter at scattering angles $> 38\arcdeg$ \citep{Mulders2012}.

\begin{figure}
\centering
\includegraphics[scale=0.3,trim=31mm 0mm 35mm 0mm,clip]{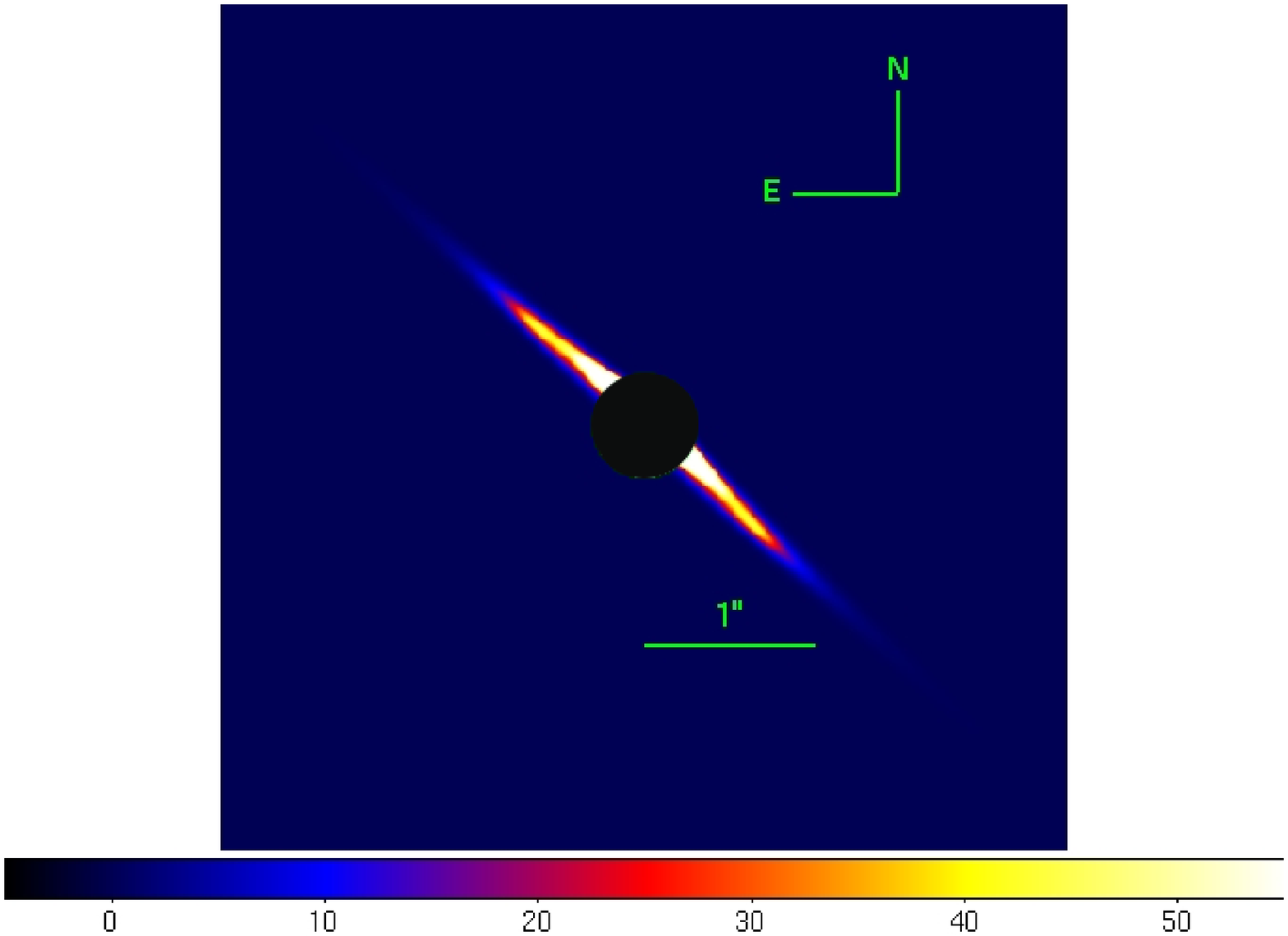}
\includegraphics[scale=0.3,trim=31mm 0mm 35mm 0mm,clip]{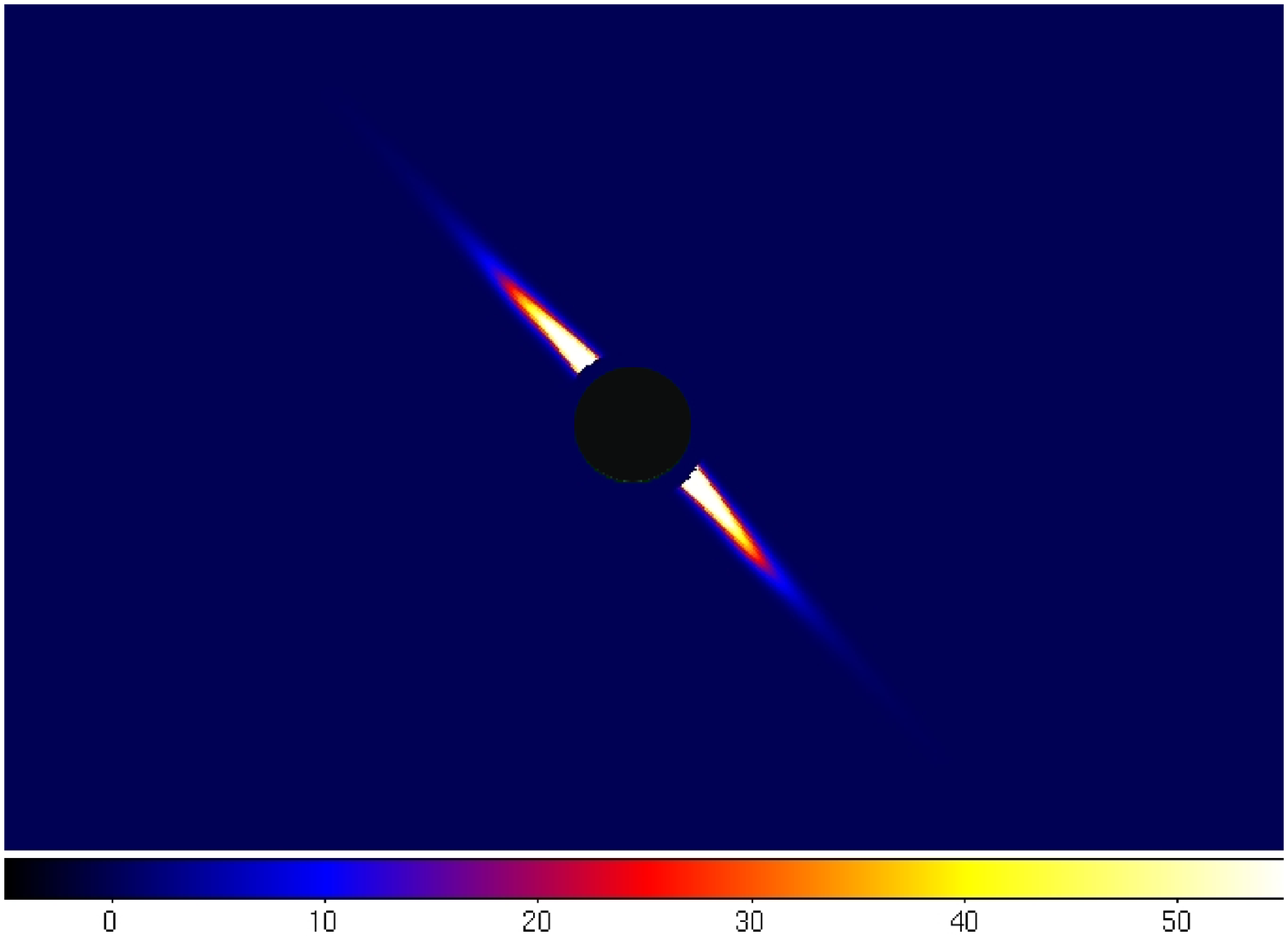}
\includegraphics[scale=0.3,trim=31mm 0mm 35mm 0mm,clip]{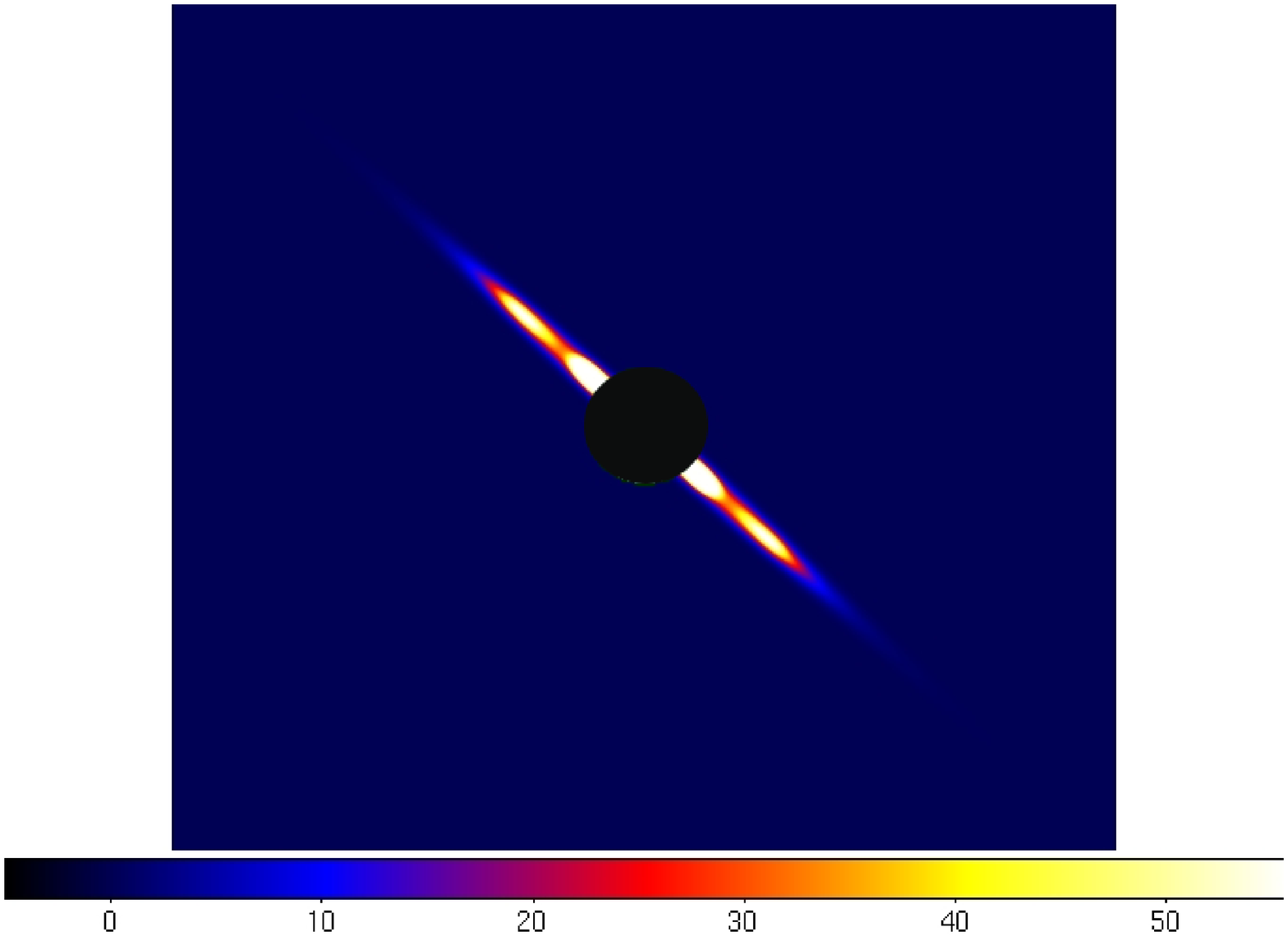}
\caption{Scattered light models incorporating different grain properties/disk geometries: 
(left) a two-component Henyey-Greenstein model with strongly forward scattering grains at small scattering 
angles but weakly scattering ones at larger angles (larger projected separations), (middle) 
a simple forward scattering grain model, (right) two isotropically scattering dust rings.  
The lefthand model best reproduces the disk SB profiles.
The units are in counts.}
\label{scatlightmod}
\end{figure}

Figure \ref{scatlightmod} displays the model that best reproduces the overall disk morphology and 
two rejected models.
We attempted several possible structures and combinations of parameters, obtaining a good fit to the
data with the model presented in Table \ref{tab:params} and shown in the left panel.  
A more rigorous exploration of parameter
space and their possible degeneracies is beyond the scope of this paper.  

To account for its warped, bow-shaped appearance, the disk must contain some highly forward 
scattering grains at $r$ $\sim$ 110 AU, which cause a brightness asymmetry between the front and back sides 
of the disk.  At low scattering angles (small projected separations) the brightness asymmetry is
more pronounced than at larger projected separations, causing a change in the midplane position
angle.  This type of ``warping'' structure is also seen in HD~15115 \citep{Debes2008,Rodigas2012}.  However,
the sharp breaks in the surface brightness at $\sim$1\arcsec\ are hard to reproduce with typical single
component Henyey-Greenstein phase functions (middle panel), requiring a flatter phase function at larger scattering
angles (left panel).  The surface brightness breaks could conceivably be reproduced by a superposition of two rings
of isotropically scattering dust (right panel), but such a configuration does not give rise to the 
disk's warped appearance.
We cannot exclude the presence of a second inner ring (i.e. at $r$ $\sim$ 35--50 AU) 
for our modeled dust phase function, especially if the dust in the disk at wider separations is slightly
less forward scattering than we have modeled.  This would 
allow the inner disk to dominate the surface brightness at small projected separations.

\begin{figure}
\plotone{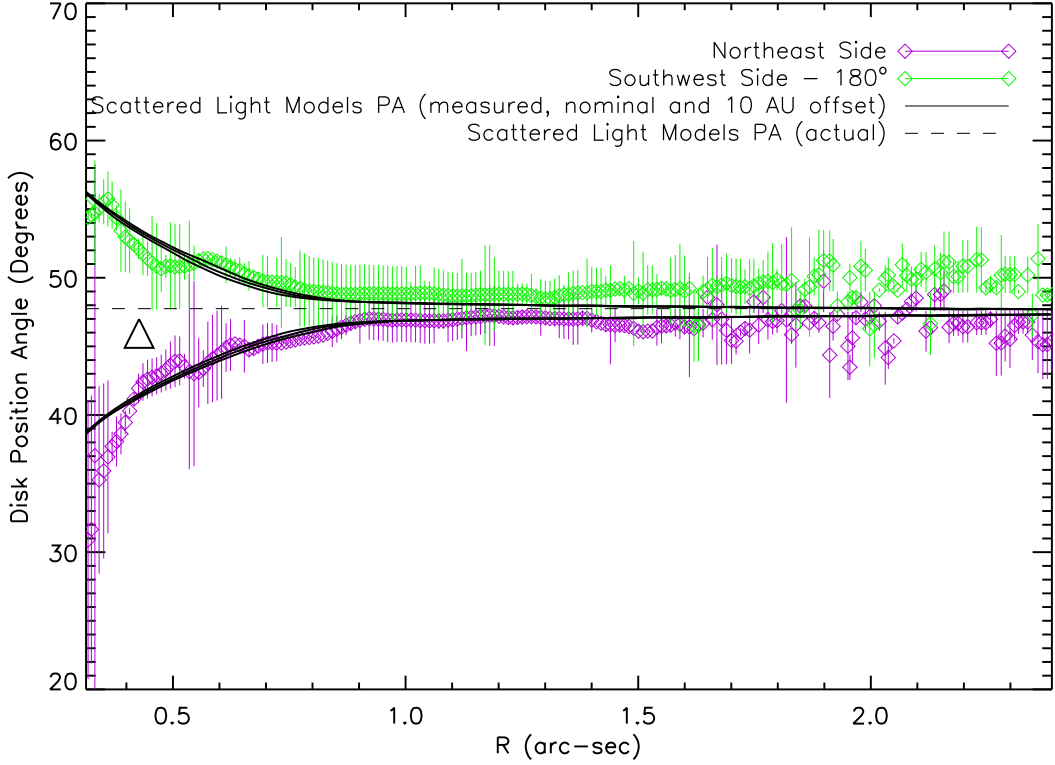}
\caption{Comparisons between the modeled and measured 
 disk position angles (right).  The triangle identifies the location of the mm brightness peak 
from \citet{Maness2008}.  We use the two-component model with the dust ring centered on the star and with 
a 10 AU offset.}
\label{scatlightcomppa}
\end{figure}

\begin{figure}
\plottwo{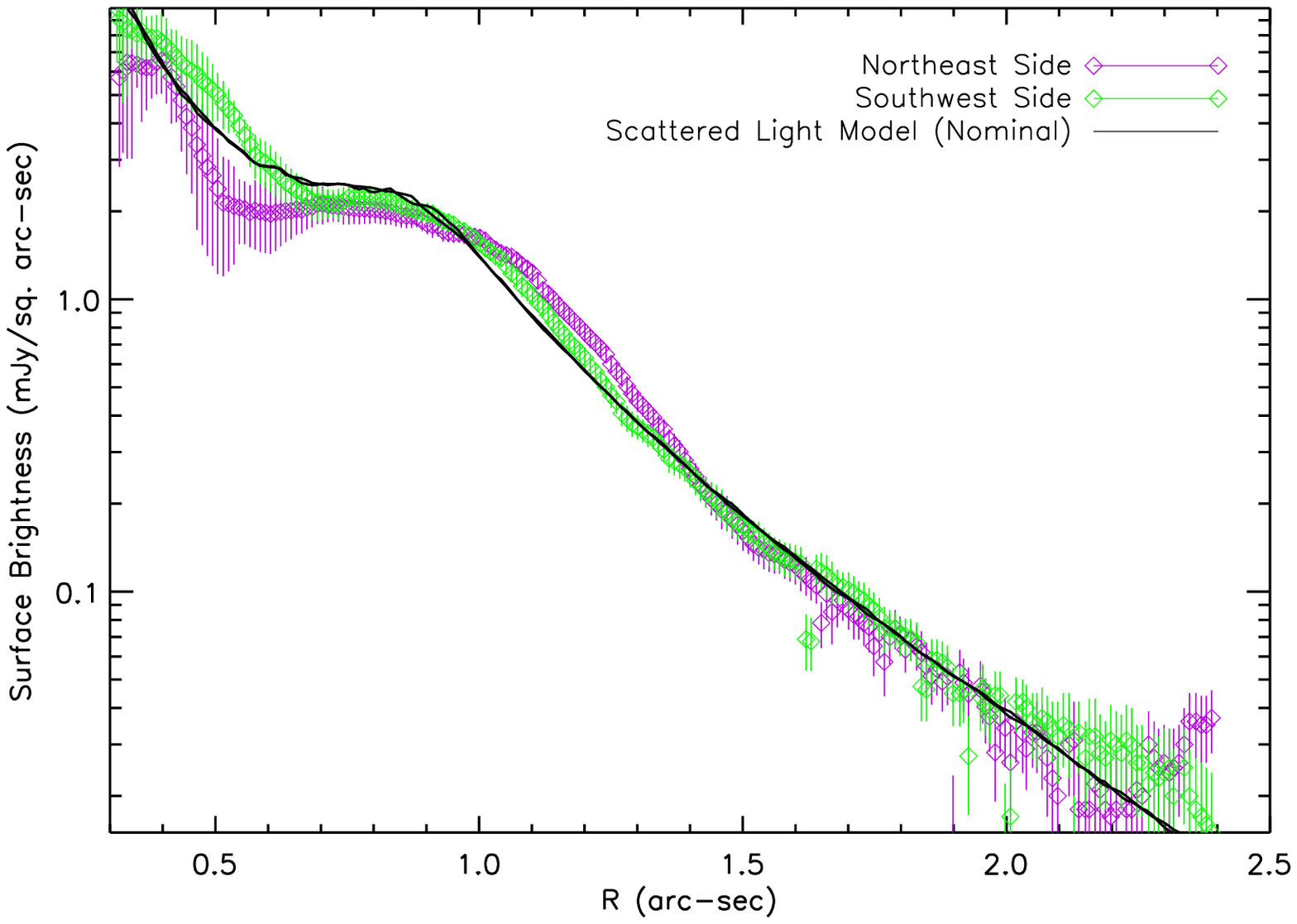}{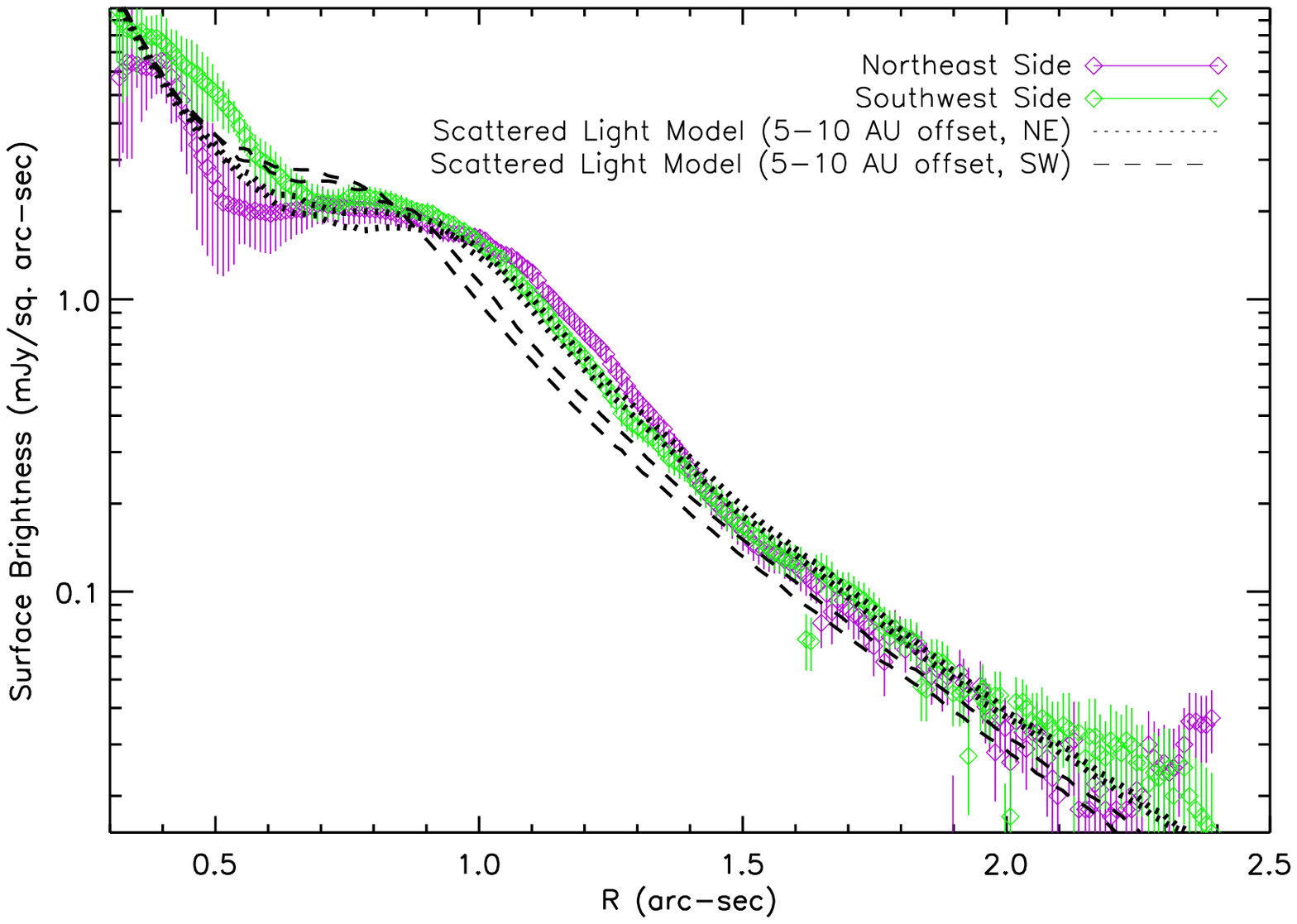}
\caption{Comparisons between the modeled and measured surface brightness profiles for a dust ring 
centered on the star (left) and one with a 5--10 AU pericenter offset (right; the SW side is 5--10 AU closer).  
Both models reproduce the wavy SB profile.  The offset makes the disk model qualitatively reproduce the differences 
in the observed SB peaks/plateaus at $r$ $<$ 1\arcsec{}, although it degrades the model's fidelity 
at $r$ $>$ 1 \arcsec{}, especially on the SW side, where it is substantially underluminous.}
\label{scatlightcompoffset}
\end{figure}

Although our two-component forward scattering model is simple, it reproduces some key disk features.
Figure \ref{scatlightcomppa} and \ref{scatlightcompoffset} (left panel) compares this model to the observed SB profile.
The model clearly succeeds in reproducing the disk warping at $r$ $<$ 1\arcsec{} and the 
surface brightness profile break/turnover at $\sim$ 110 AU.  The model grains' strong forward scattering at small angles causes the 
disk to appear very bright again at small projected separations ($r$ $<$ 0.6\arcsec{}).  This feature agrees with our 
measured SB profile (Figure \ref{scatlightcompoffset}), though taken at face value the model predicts that this brightness accelerates, whereas the measured 
profile appears to flatten at $r$ $\sim$ 0.4\arcsec{}, especially for the NE side.  Furthermore, our scattered light model 
predicts that the disk emission should be roughly axisymmetric.  However, at $r$ $<$ 0.6\arcsec{} 
the disk's SW side is significantly brighter and the profile breaks appear offset.       

The disk's asymmetric SB profile breaks could indicate an asymmetry in the disk's distance from the star as a function of 
position angle: e.g. a pericenter offset.  To investigate whether a pericenter offset better reproduces the SB profile, we 
reconstruct a scattered light model with identical grain scattering properties as our two-component model but make the 
SW side of the disk 5--10 AU closer to the star than the NE side (Figure \ref{scatlightcompoffset}).  
This model predicts that the NE side's SB profile break starts at wider separations ($r$ $\sim$ 1.1\arcsec{} 
vs. 0.9\arcsec{} for the SW side) and that the SW side should be brighter at 0.5--0.9\arcsec{}, in agreement with the 
observed SB profiles.  

The pericenter offset model's fidelity isn't perfect: the SW side is significantly underluminous and the 
exact locations of the breaks are not well reproduced.  However, its success in reproducing the 
asymmetric SB profile breaks motivates more detailed scattered light disk modeling.  In particular, all single ring 
models (regardless of any pericenter offset) predict that the disk's sides should be equally luminous at $r$ $\sim$ 0.3--0.4\arcsec{}.  
Our data indicate otherwise (the SW side is still brighter by $\sim$ 50\%), suggesting the need for further modifications 
for our scattered light model.  New, higher SNR images of 
HD 32297's disk will clarify the angular separation range over which the SW side is brighter and thus provide 
important constraints for future disk modeling.

\subsection{Spectral Energy Distribution Modeling}
For a separate but related probe of the HD 32297 disk properties, we 
modeled the disk spectral energy distribution (SED) from point-source 
photometry including newly-available 
data from the \textit{Spitzer Space Telescope}, \textit{WISE} mission, 
and \textit{AKARI} satellite \citep{Werner2004,Wright2010, Murakami2007}.  
Table \ref{phottable} lists 
the photometric data we consider.

Our methods follow those outlined in \citet{Plavchan2009}, where we 
identify the best-fit dust temperature(s), grain properties, and effective emitting areas 
incorporating a downhill simplex algorithm (``Amoeba") as described in 
\citet{Press1992}.  We consider cases where the dust radiates like a blackbody such 
that $r_{dust,AU}$ = (280K/T)$^{2}$$\times$$\sqrt{L_{star}/L_{\odot}}$ and 
where the dust's emissivity scales with the effective grain size beyond a 
critical wavelength \citep{Backman1992}, $\epsilon$ 
$\propto$ $\lambda^{-\beta}$, which can place grains of a given temperature at larger 
distances.  We solve for the disk model parameters
that minimize the fit residuals relative to the flux ($rms_{rel}$ = $\sqrt{(\sum_{i=0}^{n}\Delta_{i}/Flux_{i})^{2}/N}$).
In all cases, we assume that the grain populations have a characteristic size: for simplicity, 
we do not consider a grain size distribution.

Even though we include photometry not previously modeled, 
our fits are likely to be highly degenerate \citep[e.g.][]{Maness2008,Fitzgerald2007}.  Therefore, 
instead of identifying the single best-fit model and 95\% confidence interval, we use several
 separate SED model runs to explore more focused questions about the HD 32297 disk properties:
\begin{itemize}
\item \textbf{Model 1 -- A Single Dust Population} -- Here we assume that only one dust population contributes to the 
disk's IR-to-mm emission and allow $\beta$ to vary.  Our goal with this fit is to determine whether 
HD 32297's disk must have more than one dust population.
\item \textbf{Model 2 --  Two Dust Populations, One Dust Belt at a Fixed Location} -- Here we include two dust populations and allow 
their emissivity laws to vary but fix them to the same location.  We place the belt at 
$r$ = 85 AU, or roughly interior to where the disk surface brightness profile begins to flatten.
Our goal here is to determine whether more than one dust location is required.
\item \textbf{Models 3--5 -- Two Dust Populations/Belts, Fixed Emissivity} -- Here we identify the 
best-fitting model with two dust populations in two distinct belts, assuming either that the grains 
behave as blackbodies or have a $\lambda^{-1}$ emissivity law.  This approach follows that of 
\citet{Maness2008} and will allow us to assess whether their formalism provides a better match 
to the photometry than assuming one dust population/belt.
\item \textbf{Model 6 -- Two Dust Populations/Belts, Variable Emissivity, Fixed Location} -- Here, we fix the locations 
of the dust to the peak of the surface brightness profiles: $\sim$ 0.4\arcsec{} and $\sim$ 1\arcsec{} or 45 AU 
and 110 AU, which roughly cover the locations of the SB plateaus in the NE and SW sides.  
Our goal here is to see whether we can identify a good-fitting disk model that identifies the local maxima in the 
surface brightness profile as the locations of two separate debris belts.
\item \textbf{Models 7-9 -- Three Dust Populations/Belts, Variable Emissivity, Variable/Fixed Location} -- 
\citet{Maness2008} suggest that three separate dust populations are needed to fit the HD 32297 SED.  
In one case, we fix all the belts to be at 85 AU and incorporate a third, warmer dust population varying 
in grain size and emissivity.
In two cases, we fix the outer two dust belts at 45 AU and 110 AU, while varying grain size and emissivity and 
incorporate a third warmer dust population varying in in radius, grain size, and emissivity.
\end{itemize}

\begin{figure}
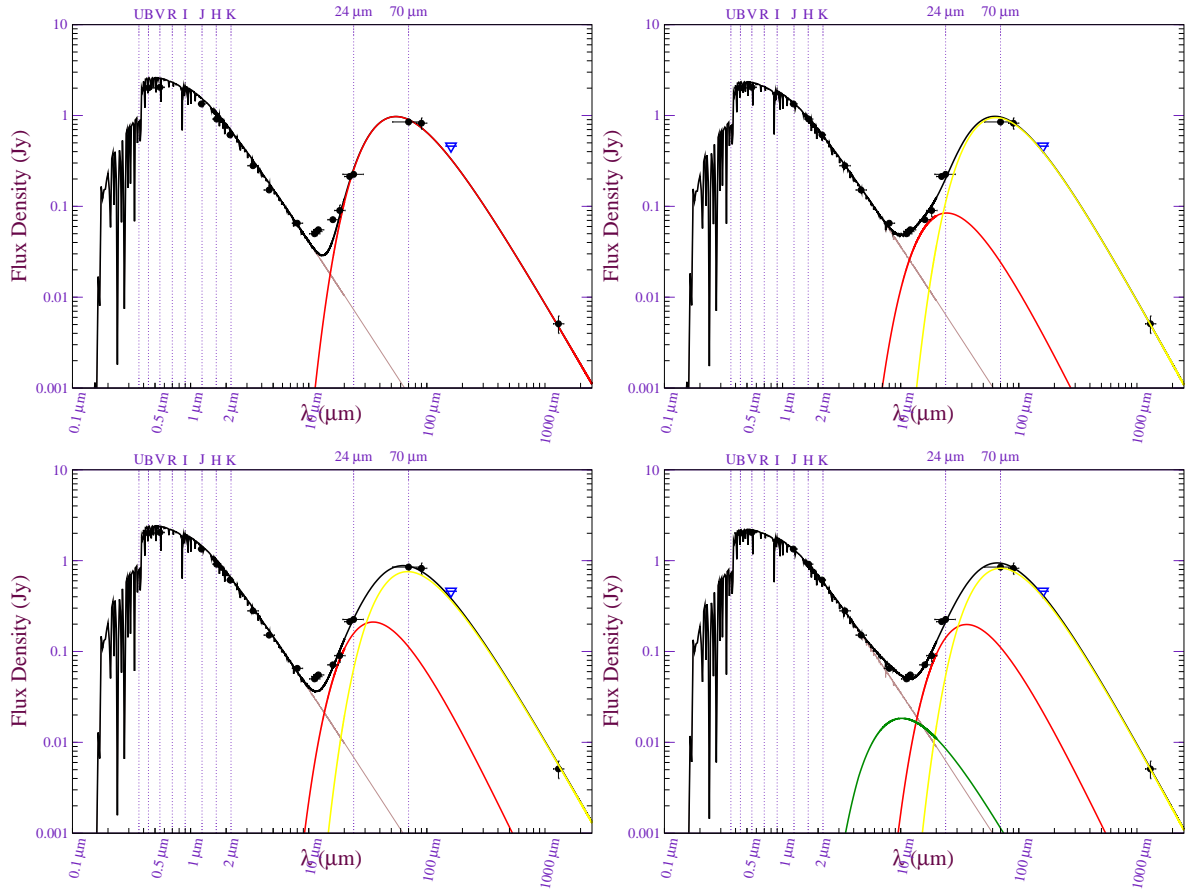

\centering
\includegraphics[scale=0.32,trim=0mm 0mm 0mm 10mm,clip]{sedmodel1.eps}
\includegraphics[scale=0.32,trim=0mm 0mm 0mm 10mm,clip]{sedmodel2.eps}
\\
\includegraphics[scale=0.32,trim=0mm 0mm 0mm 10mm,clip]{sedmodel6.eps}
\includegraphics[scale=0.32,trim=0mm 0mm 0mm 10mm,clip]{sedmodel9.eps}
\caption{SED model fits to the HD 32297 photometric data listed in Table \ref{phottable}: 
Model 1 (top-left), Model 2 (top-right), 
Model 6 (bottom-left), and Model 9 (bottom-right)}.
\label{sedfits}
\end{figure}

Figure \ref{sedfits} displays some of our SED modeling results, which are reported in 
Table \ref{sedtable}.  For all best-fit models,
the stellar effective temperature is $T_{\star}$ $\sim$ 7870-7890 K and the star has 
little extinction ($A_{V}$ $<$ 0.03).  Regardless of our assumed particle emissivity law, 
a single dust population model (Model 1, top-left panel) poorly reproduces the SED, 
as was found previously \citep{Fitzgerald2007}, significantly underpredicting the flux density 
at 8--22 $\mu m$.   Formally, having two or three dust populations in one belt at 85 AU 
(i.e. top-right panel) significantly 
improves the fit ($rms_{rel}$ = 0.05--0.08).  However, the required grain sizes are too small 
($\sim$ 4 nm--0.2 $\mu m$) to be realistic.  Thus, the dust likely arises from more than 
one single-temperature dust population with given grain size and emissivity power law.

Fits assuming blackbody grains but incorporating two dust components 
at different locations significantly 
improve the fit ($rms_{rel}$ $\sim$ 0.16; top-right) compared to a single 
dust component model.
  The fits imply that the dust responsible for 10--1000 $\mu m$ 
emission is at 1.2--22 AU or $r$ $\sim$ 0.01--0.2\arcsec{}, but \citet{Fitzgerald2007}, 
\citet{Moerchen2007}, and \citet{Maness2008} show that the disk emission instead originates on 
scales more comparable to $\approx$ 30--100 AU.  
The broad range of orbital radii is indicative of the significant model degeneracies involved.
Assuming that at least one of the dust components has an emissivity of $\beta$ = 1 
yields fits with larger grains (0.3--3 $\mu m$) but worsens the 
fit ($rms_{rel}$ $\sim$ 0.17--0.21).  Thus, if the dust disk geometries approximate thin, isothermal 
rings and consistent of grains dominated by a single, characteristic size, the grains likely have emissivity power laws
 somewhere between 0 and 1.  

The bottom panels of Figure \ref{sedfits} show that it might be possible for 
the debris rings to reside at locations equal to the scattered light brightness peaks ($r$ $\approx$ 45 AU, 110 AU) 
and reproduce the IR to mm disk emission.  A two dust population model (Model 6; bottom-left) reproduces the 
SED well at 1--8 $\mu m$ and 17--1300 $\mu m$ but underpredicts the disk emission at $\sim$ 12 $\mu m$ 
by $\sim$ 50\%.  Starting with the Model 6 results and incorporating a third dust population, we 
reproduce the SED if the dust is at $\sim$ 14 AU (171 K), although the emissivity power law for this 
population is unphysical ($\beta$ $\sim$ 7.3)\footnote{This model run finds $\beta$ $\sim$ 7.3 as the ``best-fit" 
model because the Amoeba code treats $\beta$ as a true free parameter, regardless of whether the value is 
physical, and exploits a gap in SED coverage at $\sim$ 10 $\mu m$ to achieve a better fit.} 
and the grain sizes for the coldest dust component are 
too small (Model 8).  However, allowing the emissivity law and grain sizes for all three 
components to vary yields an excellent fit to the data (RMS $\sim$ 0.06).  The emissivity laws and 
grain sizes for this model (Model 9; bottom-right) are reasonable ($\beta$ = 0.37--0.77; 
$a$ = 0.20--0.70 $\mu m$).  Here, the third component is at $\sim$ 1 AU with a temperature of $\approx$ 430 K and 
is responsible for the weak 8 $\mu m$ excess and substantially contributes to the disk emission at 12--17 $\mu m$.  
Thus, although it is not \textit{necessary}, it is at least \textit{possible} to identify the 
surface brightness profile peaks at 45 AU and 110 AU with separate thin dust rings responsible for 
broadband emission at 8--1300 $\mu m$ \textbf{provided} that there exists an unresolved, warmer dust 
component located interior to 45 AU.  

We emphasize that the model fits we report in Table \ref{sedtable} are but some of many possibile fits, the 
range of which is further limited by our input assumptions about the disk.
For example, because our model is set up only to consider infinitely thin, isothermal 
rings, it is unclear whether the warmer dust emission we identify originates in a separate ring or rather an annulus, the 
outer edge of which we see at wider separations.  Given the extreme number of model degeneracies, 
the most we can say is that 1) the debris emission must originate from more than one population 
at 2) multiple locations and 3) among the many possibile configurations, the dust populations may coincide 
with the surface brightness peaks.  To make further progress, we need high spatial resolution, high SNR imaging 
of the disk at a wider range of wavelengths to derive much more stringent 
constraints on the dust sizes/scattering properties and location(s) of the dust population(s) 
\citep[e.g.][]{Debes2009,Rodigas2012}.
\section{Comparisons to Previous HD 32297 Results}
From optical coronagraphic imaging, \citet{Kalas2005a} find evidence for warping in the HD 32297's disk, 
especially for the NE side, where the disk at $r$ $>$ 500--600 AU appears swept back compared to midplane regions at smaller 
angular separations.  They attribute the warp to ISM sculpting of the disk as it moves south through 
ISM material.  The warped structure we resolve curves in the opposite direction from that expected due to 
ISM sculpting \citep[see images for HD 32297/15115/61005 from][]{Debes2009} and does so for both sides of the disk.  
Furthermore, ISM sculpting should be prominent only at larger separations in our images, where the 
disk is weakly detected ($r$ $>$ 1.5\arcsec{}), because the grains need time to become entrained in the 
ISM flow \citep[e.g.][]{Debes2009}. 
However, optical and near-IR imaging probe two very different characteristic grain populations, and the size scales over 
which we resolve the disk do not overlap.  Thus, our results are not in conflict.  Rather, ISM sculpting may explain 
the optical image but does not explain the near-IR image.

We confirm and clarify three major claims from previous near-IR imaging \citep{Schneider2005,Mawet2009,Debes2009}.  
First, we verify the brightness asymmetry found by \citeauthor{Schneider2005} and \citet{Mawet2009} 
at small ($r$ $\sim$ 45--55 AU) separations and show that it persists down to $r$ $\sim$ 35 AU, albeit 
at a lower statistical significance.  
Second, we recover a break/deviation in the SB power law at $r$ $\sim$ 100 AU found by \citet{Debes2009}.  
Our new results show that the SB profile at smaller separations does not follow an extension of the 
power law describing the disk at larger separations: instead, it appears wavy.
Finally, \citet{Mawet2009} 
claim that the NE side of the disk emission is truncated at $r$ $\sim$ 0.6\arcsec{}.
We find that the NE SB profile flattens (not brightens) from $r$ $\sim$ 1\arcsec{} to $r$ $\sim$ 0.5\arcsec{}.  
Assuming that their coronagraph attenuates some flux at separations slighter greater than their 
inner working angle (0.4\arcsec{}), most of the NE side emission will be hidden from their view.  In contrast, 
the SW side clearly begins to brighten at wider separations ($r$ $\sim$ 0.5--0.6\arcsec{}) visible by 
\citet{Mawet2009}.  Thus, our results agree.   

Thermal IR imaging from \citet{Fitzgerald2007} and \citet{Moerchen2007} shows that the HD 32297 disk has an inner clearing 
devoid of grains slightly larger than those we can probe with our data.  \citeauthor{Fitzgerald2007} find that the 
disk exhibits a bilobed structure and most of its 11.2 $\mu m$ emission originates from $r$ $\sim$ 0.5--0.6\arcsec{}, characteristic 
of a dust ring with an inner edge at $\approx$ 60 AU.  
\citeauthor{Moerchen2007} resolve the disk at 12 $\mu m$ and 18 $\mu m$ out to slightly wider 
separations ($r$ $<$ 1.3\arcsec{}).  Based on the disk's brightness temperature, they likewise 
find evidence for an inner clearing, albeit one that is slightly larger ($r$ $\sim$ 80 AU).
The plateaus in our SB profiles could identify the inner boundaries of dust 
populations \citep[e.g.][]{Rodigas2012}.  Under this interpretation, our image is consistent with 
a dust belt truncated at $r$ $\sim$ 1\arcsec{} ($\approx$ 110 AU) and a second one 
at $\approx$ 0.4--0.6\arcsec{} (45--70 AU).  Because these authors did not analyze their data in 
the same way it is difficult to compare their results between each other and their combined results 
against ours.  Still, their results and ours are qualitatively consistent with the HD 32297 disk 
having at least one dust population truncated at small separations. 

In agreement with \citet{Maness2008}'s mm study, we find evidence for at least two dust grain populations, 
responsible for the mid-IR excess emission and far-IR/submm excess, respectively \citep[see also][]{Moerchen2007}.  They are able to 
fit the SED from 1 to 100 $\mu m$ but not 1300 $\mu m$, a discrepancy they explain by adding a third, cold dust 
population.  Our modeling generally fits the SED from 17 $\mu m$ to 1300 $\mu m$ but slightly underpredicts 
emission at 12 $\mu m$, a discrepancy we can solve if there exists an additional, unresolved 
warm dust population.  Our different results are likely a byproduct of our modeling assumptions: they allow 
the dust populations to be spatially extended but fix the particle emissivity law, whereas we assume the dust 
is confined in isothermal rings but allow the emissivity law to vary.  

As noted in \S 3, the peak of HD 32297's mm emission from \citeauthor{Maness2008} 
lies close to the bright inner disk region 
on the SW side.  While formally the position angle ($\approx$ 46$^{\circ}$) of the mm emission is offset by $\approx$ 10 degrees, it 
tracks the disk's position angle at wider separations much better.  The disk's warped appearance at $r$ $<$ 1\arcsec{} is due to its 
strong forward scattering at small angles.  If the disk's grains instead isotropically scattered starlight, its $K_{s}$ emission would lie almost directly 
on top of the mm peak.  Thus, we identify the mm and near-IR brightness asymmetry as originating from the same location.  
Assuming the grains responsible for both the mm and near-IR emission are likely the result of collisions, they may
 trace the same parent population of planetesimals. 
\subsection{Summary and Future Work}
Using Keck/NIRC2 $K_{s}$ coronagraphic imaging, we resolve the HD 32297 disk at a high signal-to-noise from $r$ = 0.3\arcsec{} 
to $r$ = 2.5\arcsec{}.  We determine basic disk properties (SB, FWHM, position angle), compare our image to disk models 
with a range of (grain) scattering properties, and model newly-available, broadband photometry to provide a complementary 
investigation of HD 32297's circumstellar environment.  We obtain the following major results:
\begin{itemize}
\item \textbf{1.} We discover that HD 32297's debris disk exhibits a prominent warped or ``bow"-shaped structure interior to 
$r$ $\sim$ 1 \arcsec{} ($\sim$ 110 AU).
\item \textbf{2.} Our new data clarifies the disk's surface brightness profile at small separations.  
We find that it has a ``wavy" profile interior to $r$ $\sim$ 110 AU with a plateau 
extending to $r$ $\sim$ 0.5--0.7\arcsec{} (55--80 AU) before the disk brightens by factors of 3--4 at smaller separations.
\item \textbf{3.} The disk exhibits significant asymmetries between the two sides.  
  The SW side is brighter at $r$ $\sim$ 0.3--0.6\arcsec{} by 50--150\%, with the most statistical differences being at 
$r$ $\sim$ 0.5--0.6\arcsec{}.  These separations are roughly where the disk's mm emission peaks and are 
consistent with previous results \citep[e.g.][]{Schneider2005}.  
  Our new analysis identifies new asymmetries, revealing that the locations of the peaks/plateaus in surface brightness are likely
 shifted between the two sides, consistent with non-azimuthally symmetric structure.
\item \textbf{4.} A disk model with a flat phase function and strongly forward scattering grains 
where the dust ring is centered on the star reproduces the ``bow" structure,  marginally reproduces the ``wavy" SB profile and fails to reproduce the NE/SW asymmetries.  
A disk model with a 5--10 AU pericenter offset reproduces the asymmetric SB profile breaks interior to $r$ $<$ 1\arcsec{}, although 
its brightness asymmetry is limited to 0.5--0.9\arcsec{} and its match to the SB at wider separations is far poorer.
Thus, dust scattering plays a critical role in explaining key observed disk properties, but it is unclear whether it explains 
\textit{all} of the disk's properties we identify.
\item \textbf{5.} HD 32297 must be surrounded by more than one dust population likely arising 
from different locations in the disk.  Although these populations need not be identifiable from our image, 
we can fit the disk SED by placing dust populations at the locations of the SB peaks from our image provided that 
there exists additional warm dust that we cannot yet resolve. 

\item \textbf{6.} The disk's brightness peak at $r$ $\approx$ 0.4\arcsec{} coincides with the peak mm emission 
\citep{Maness2008}.   If the grains responsible for both peaks are the result of collisions, they may trace the 
same parent population of planetesimals.  
\end{itemize}

In summary, the HD 32297 disk appearance is broadly shaped by ISM interactions \citep{Kalas2005a,Debes2009} at wide separations and 
by its grain scattering properties at small separations (this work).  However, it is unclear whether either of these features by 
themselves explain the disk's SB profile and (especially) the disk's brightness asymmetry.   Furthermore, SED modeling provides evidence for 
multiple dust populations, possibly multiple belts.
To explain the disk's SB profile/multiple dust populations and (especially) brightness asymmetry, we may need to
 invoke other mechanisms.

Planets can sculpt dust into debris rings
\citep[e.g.][]{Kalas2005b,Quillen2006,Kalas2008}.  Furthermore, planets 
can trap dust into resonant structures, which can appear as bright, overdense regions like the 
brightness peak imaged here and in the mm at $r$ $\sim$ 0.4\arcsec{}
 \citep[e.g.][]{Liou1999,Kuchner2003,Stark2008}.  As argued by \citet{Wyatt2006} and 
\citet{Maness2008}, the detectability of the resonant structure may be wavelength dependent.    
Large grains producing mm emission are fragments of the colliding planetesimals in resonance, but 
cannot be rapidly removed by radiative forces.  Thus they can 
trace the parent body resonant structure.  Small, (sub)-micron sized grains likewise may trace resonant structure since 
they are preferentially produced in the most collisionally active, highest density regions and are otherwise quickly 
removed by radiation pressure.  Grains with intermediate sizes producing emission at intermediate wavelengths (e.g. thermal IR) 
can be pushed out of resonance by radiation pressure/PR drag but are not small enough to be rapidly removed 
from the system (Wyatt et al. 2006; though see Kuchner and Stark 2010).  
Thus, it is \textit{possible} that our near-IR image, when combined with the mm image, identifies 
planet-induced structure.

Further near-IR and (sub)mm imaging is required to verify whether the SB profiles and brightness asymmetries 
are bona fide evidence for an embedded planet.  The current state-of-the-art near-to-mid IR high-contrast 
imaging facilities like the \textit{Large Binocular Telescope} have already shown 
great promise for resolving disk's like HD 32297's in scattered light and imaging self-luminous 
gas giant planets \citep{Rodigas2012,Skemer2012} and upcoming planet imagers like \textit{SCExAO} on Subaru, \textit{GPI} on Gemini-South, 
and \textit{SPHERE} on the VLT will be even more capable \citep{Esposito2011,Martinache2009,Macintosh2008,Beuzit2008}.
 Moreover, the thermal IR is well suited for imaging exoplanets with a wide range of 
ages \citep[e.g.][]{Marois2010,Currie2011b,Rodigas2011}, especially for stars like HD 32297 whose 
bright, edge-on debris disk degrades planet sensitivity limits in the near-IR.  Imaging with 
the \textit{Atacama Large Millimeter Array} (ALMA) can resolve the HD 32297 debris disk up to factor of 
$\sim$ 50 better than the CARMA observations reported by \citeauthor{Maness2008} and thus will provide a far 
better probe of any planet-induced structure in the HD 32297 debris disk.

\acknowledgements 
We thank Richard Walters and Gregory Wirth for valuable help in setting up these observations, which were conducted remotely 
from the California Institute of Technology and to Randy Campbell for indispensible NIRC2 support in 
helping the observer (TC) conduct his observations efficiently.  
Karl Stapelfeldt, Scott Kenyon, and Margaret Moerchen provided very useful discussions and the anonymous referee provided helpful 
comments.
TC is supported by a NASA Postdoctoral Fellowship.
The data presented herein were obtained at the W.M. Keck Observatory, which is 
operated as a scientific partnership among the California Institute of Technology, the 
University of California and the National Aeronautics and Space Administration. The Observatory was 
made possible by the generous financial support of the W.M. Keck Foundation. 
We acknowledge the significant cultural role and reverence
that the Mauna Kea summit has always had within the
indigenous Hawaiian community. We are grateful to be able to conduct
observations from this mountain.
This research has made use of the NASA/ IPAC Infrared Science Archive, which is operated by 
the Jet Propulsion Laboratory, California Institute of Technology, under contract with the National Aeronautics and Space Administration.
{}
\begin{deluxetable}{llllllllllllll}
\tablecolumns{4}
\tablecaption{Surface Brightness Profile Power Law Indices}
\scriptsize
\tablehead{{Angular Separation (\arcsec{})} & {a,b(NE)} & {a,b(SW)}}
\startdata
Single Power Law\\
$r$ = 1--2.5 & 2.13, -6.01& 1.72, -5.49\\
\\
Broken Power Law\\
$r$ = 1.1--1.4 & 2.23,-6.19 & --\\
$r$ = 1--1.6 & -- & 1.78, -5.71\\
\\
$r$ = 1.4--2.5 &1.31,-5.13 & --\\
$r$ = 1.65--2.5 &-- & 1.66,-5.33\\
\enddata
\tablecomments{ Surface brightness profiles for the NE and SW sides determined by a Levenberg-Marquardt 
least-squares fit to the data.  We assume a simple power law functional form of $f(X)$ = $a$X$^{b}$.  Power laws 
in agreement with the data at the 95\% confidence limit include values for $a$ that vary by $\sim$ 10\% and 
values for $b$ that vary by $\sim$ 0.5 dex.} \label{plawtable}
\end{deluxetable}

\begin{deluxetable}{cc}
\tablecolumns{2}
\tablewidth{0pt}
\tablecaption{\label{tab:params}}
\tablehead{\multicolumn{2}{c}{HD~32297 Scattered Light Model Parameters}
}
\startdata
$r_{\rm o}$ & 110~AU \\
$\sigma_{\rm r}$ & 13~AU \\
FWHM$_{\rm r}$ & 30~AU \\
$\sigma_{\rm z}$ & 2~AU \\
FWHM$_{\rm z}$ & 5~AU \\
$r_{\rm break}$ & 110~AU \\
$g_1$ & 0.96 \\
$a_1$ & 0.9 \\
$g_2$ & -0.1 \\
$a_2$ & 0.1 \\
$\beta$ & 3.4 \\
$i$ & 2$\arcdeg$ \\
Position Angle & 47.5$\arcdeg$
\enddata
\end{deluxetable}

\begin{deluxetable}{llllllllllllll}
\tablecolumns{4}
\tablecaption{HD 32297 Photometric Data}
\scriptsize
\tablehead{{Filter/Wavelength} & {Flux (mJy)} & {$\sigma$Flux (mJy)} &  Source}
\startdata
$B$/0.4380 & 1836.74 & 28.75 & TYCHO-II(trans) \\
$V$/0.545 & 2012.91 & 24.10 & TYCHO-II(trans) \\
$J$/1.235 & 1341.80 & 34.96 & 2MASS \\
$H$/1.662 & 913.48 & 50.58 & 2MASS \\
$K_{s}$2.159 & 611.41 & 11.95 & 2MASS \\
3.37 & 280.23 & 8.21 & WISE \\
4.62 & 151.28 & 3.78 & WISE \\
8 & 65.18 & 2.75 & IRSep \\
11.2 & 49.9 & 2.1 & Fitzgerald et al. (2007) \\
11.66 & 53 & 5.3 & Moerchen et al. (2007) \\
12.08 & 55.25 & 1.20 & WISE \\
16 & 71.28 & 2.53 & IRSep \\
18.3 & 90 & 13.5 & Moerchen et al. (2007) \\
22.19 & 212.99 & 5.55 & WISE \\
24 & 225.2 & 4.82 & IRSep \\
70 & 850 & 60 & IRSep \\
90 & 823.2 & 116 & AKARI \\
160 & $<$460 &  & IRSep \\
1300 & 5.1 & 1.1 & CARMA/Maness et al. (2008) \\
\enddata
\tablecomments{ \textit{TYCHO-II(trans)} refers to TYCHO-II catalog data transformed into the standard Johnson-Cousins photometric 
system.  \textit{IRSep} refers to the IRS enhanced products dataset as queried from the NASA/IPAC Infrared Science Archive.}
\label{phottable}
\end{deluxetable}

\begin{deluxetable}{ccccccccccc}
\tablecolumns{11}
\tablecaption{HD 32297 SED Modeling Results}
\scriptsize
\tablehead{{Model ID} & {RMS}&{$R_{dust,1}$ (AU)} & {$R_{dust,2}$} 
& {$R_{dust,3}$}&{$\beta_{1}$}&{$\beta_{2}$}&{$\beta_{3}$} &
{$a_{1}$ ($\mu m$)} & {$a_{2}$}&{$a_{3}$}}
\startdata
1 & 0.23& 85 & -- & --& 0.94 & -- & -- & 0.095 & -- & --\\
2 & 0.08& \textbf{85} & \textbf{85} & -- & 0.72 & 0.39 & -- & 0.004 & 0.197 & --\\
3 & 0.16& 1.21& 21.66&--& $\textbf{0}$ & $\textbf{0}$&--&0.39&0.41&--\\
4 & 0.17& 55.64& 527.56&--& $\textbf{1}$ & $\textbf{1}$&--&0.30&0.99&--\\
5 & 0.21& 15.60& 97.50&--& $\textbf{0}$ & $\textbf{1}$&--&0.91&2.86&--\\
6 & 0.15& $\textbf{45}$ &$\textbf{110}$&--&0.80&0.31&--&0.39&0.025\\
7 & 0.05& \textbf{85}&\textbf{85} & \textbf{85}&0.84&0.61&0.45&0.006&0.024&0.725\\
8 & 0.09& 14& \textbf{45} & \textbf{110}&7.27&0.80&0.31&1.59 & 0.39 & 0.03\\
9 & 0.06& 1.1 & \textbf{45} & \textbf{110}& 0.37 & 0.77 & 0.43 &0.70 & 0.37 & 0.20\\ 
\enddata
\tablecomments{
RMS refers to the fit residuals relative to the flux $rms_{rel}$ = $\sqrt{(\sum_{i=0}^{n}\Delta_{i}/Flux_{i})^{2}/N}$, where 
$N$ is the number of flux density measurements (18).
 $R_{dust}$ refers to the dust ring's stellocentric distance, $\beta$ is the particle emissivity power law, 
and $a$ is the grain size in microns, where 
2$\pi$$a$ = $\lambda_{o}$ \citep[see][]{Backman1992}.  
Values in bold are fixed for a given model run, whereas others are `fitted' values.}
\label{sedtable}
\end{deluxetable}




\begin{figure}
\centering
\includegraphics[scale=0.47,trim=31mm 0mm 35mm 0mm,clip]{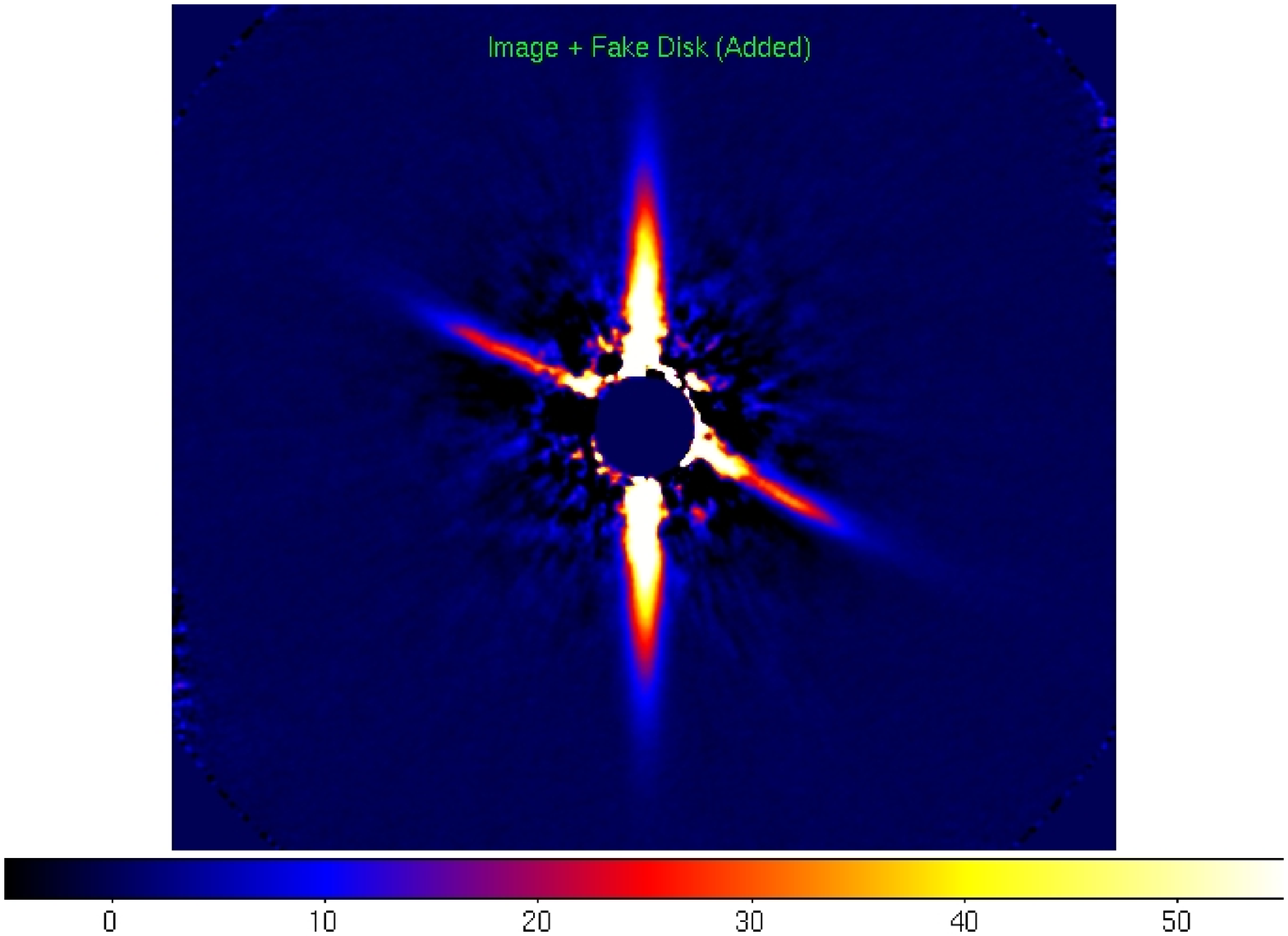}
\includegraphics[scale=0.47,trim=31mm 0mm 35mm 0mm,clip]{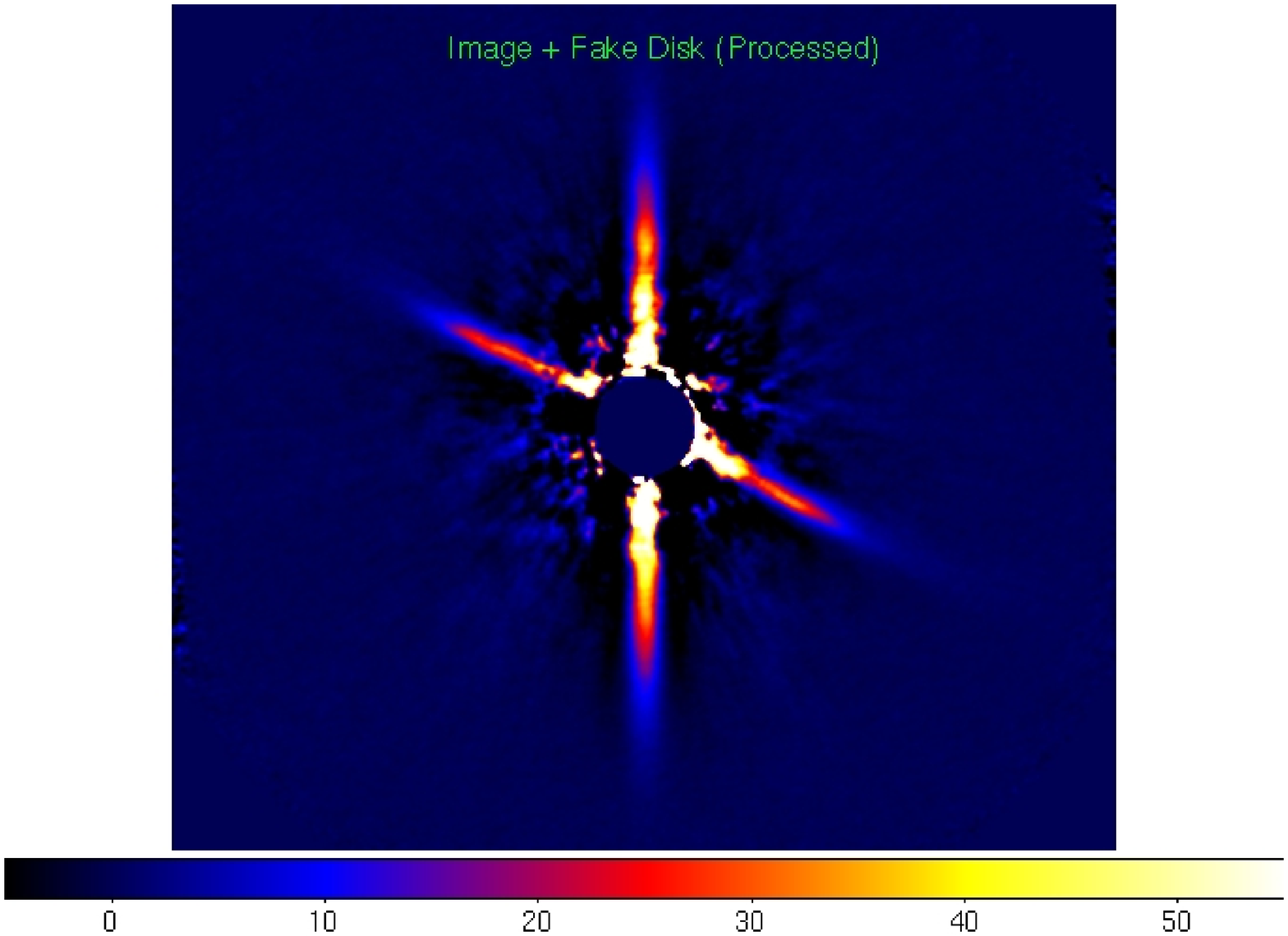} 
\caption{Images used to model biasing from LOCI processing.  (Left) Our final image with a fake disk 
added.  (Right) Our final image where we add the fake disk to each registered image and process the 
set of images with our pipeline.  In both cases, we rotate the image to the parallactic angle of the first 
image in the sequence (PA $\sim$ -15.06$^{\circ}$), not to true north as we do in Figure \ref{images1}.} 
\label{fakedisks}
\end{figure}

\begin{figure}
\plottwo{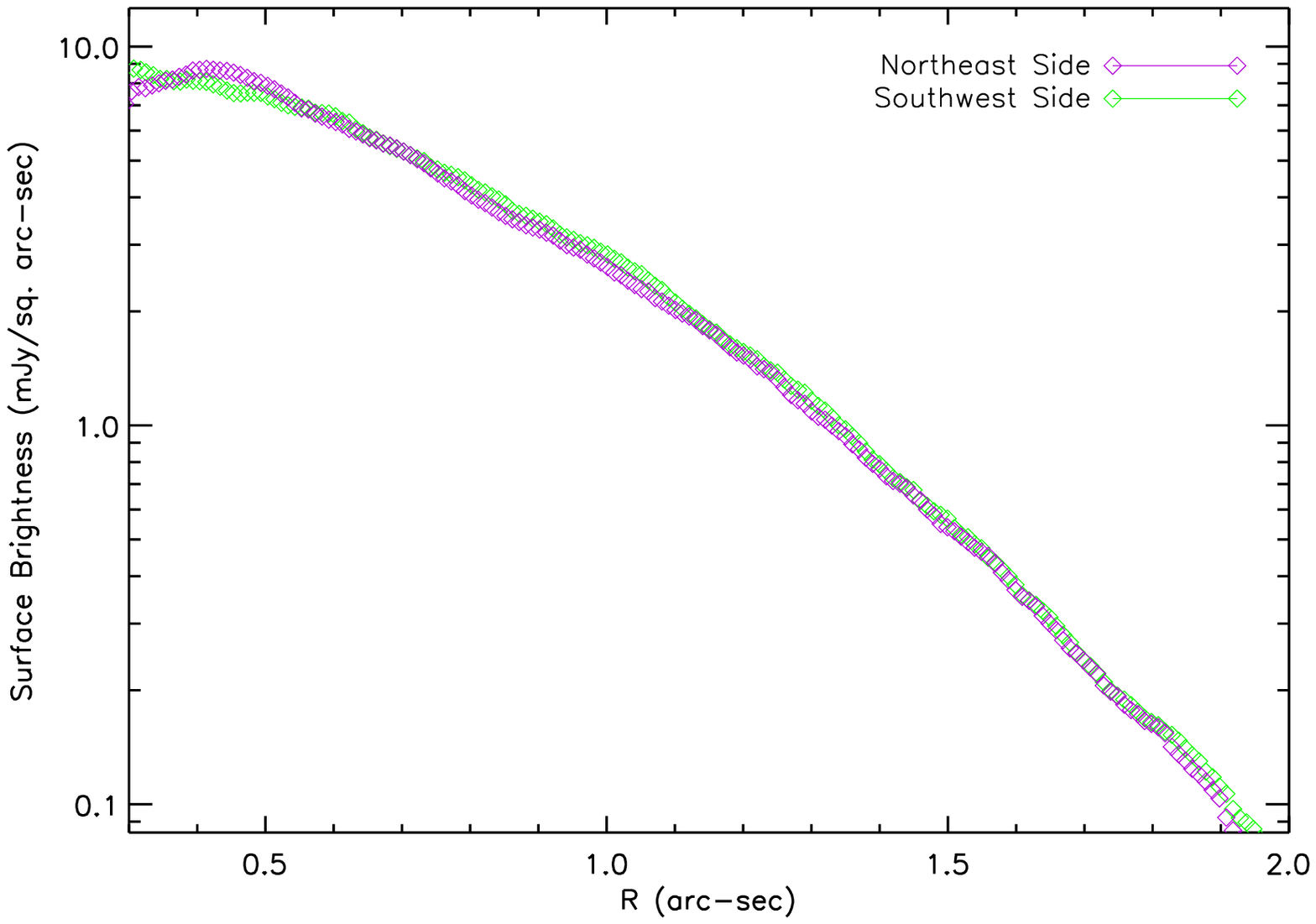}{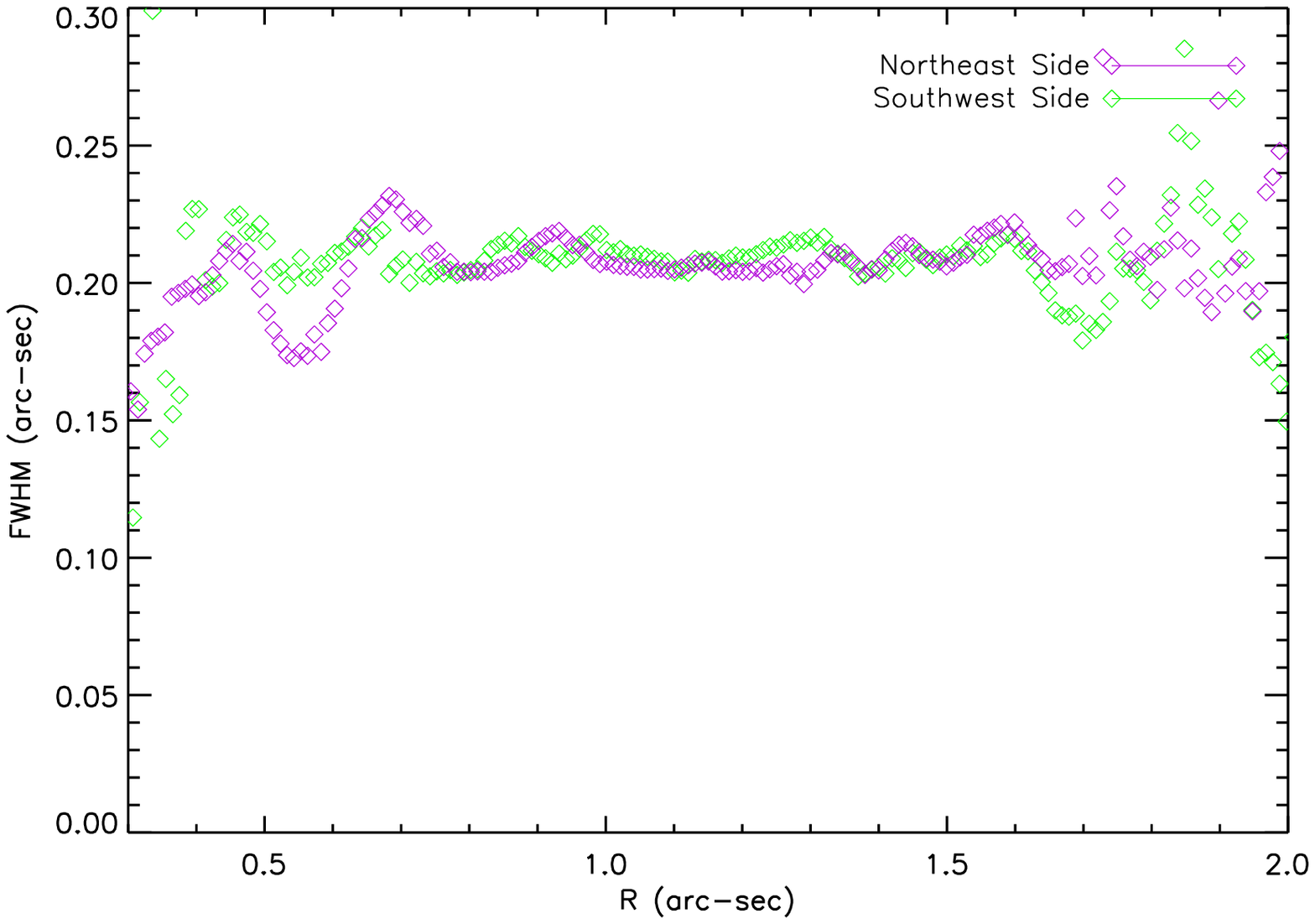}
\plottwo{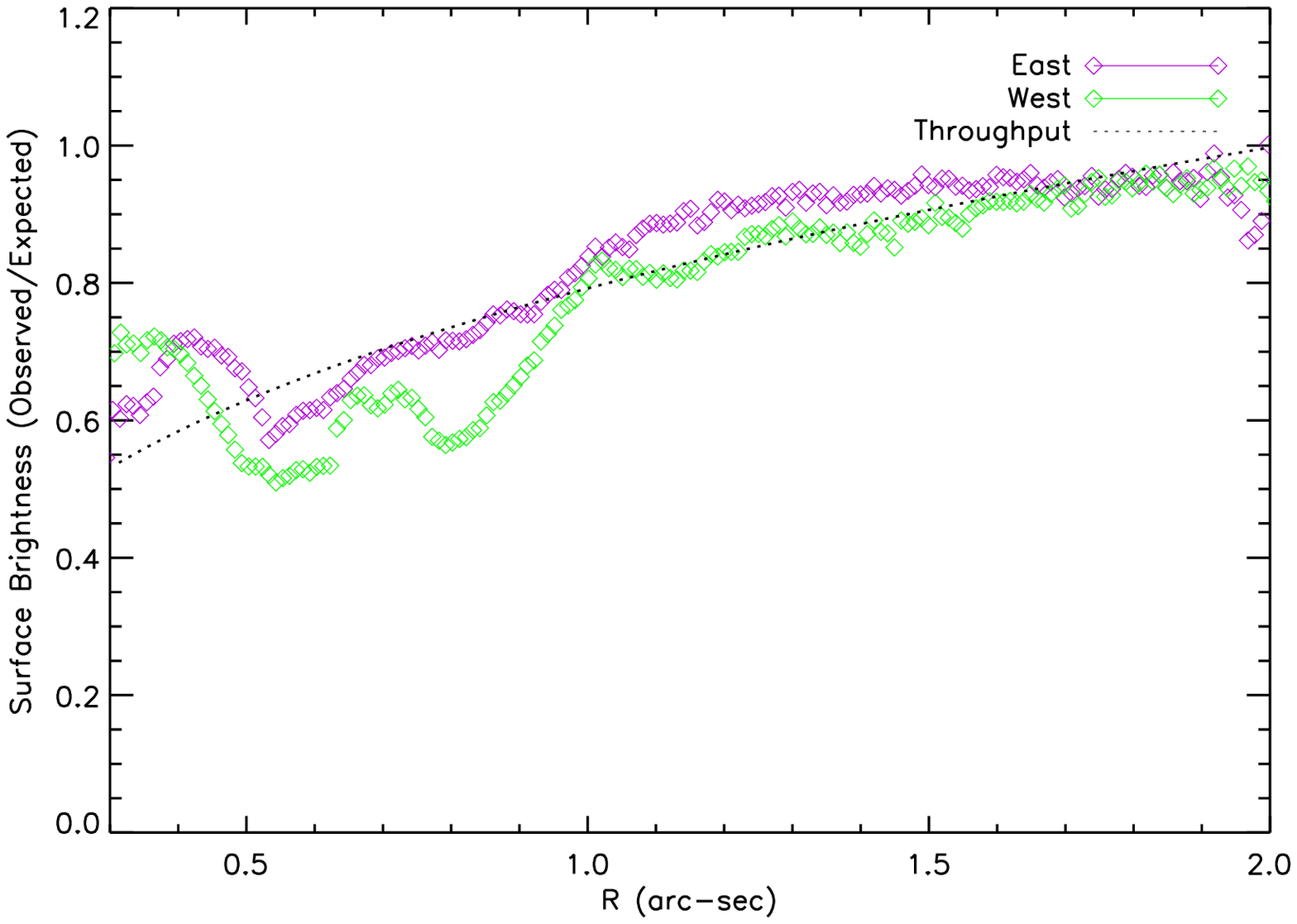}{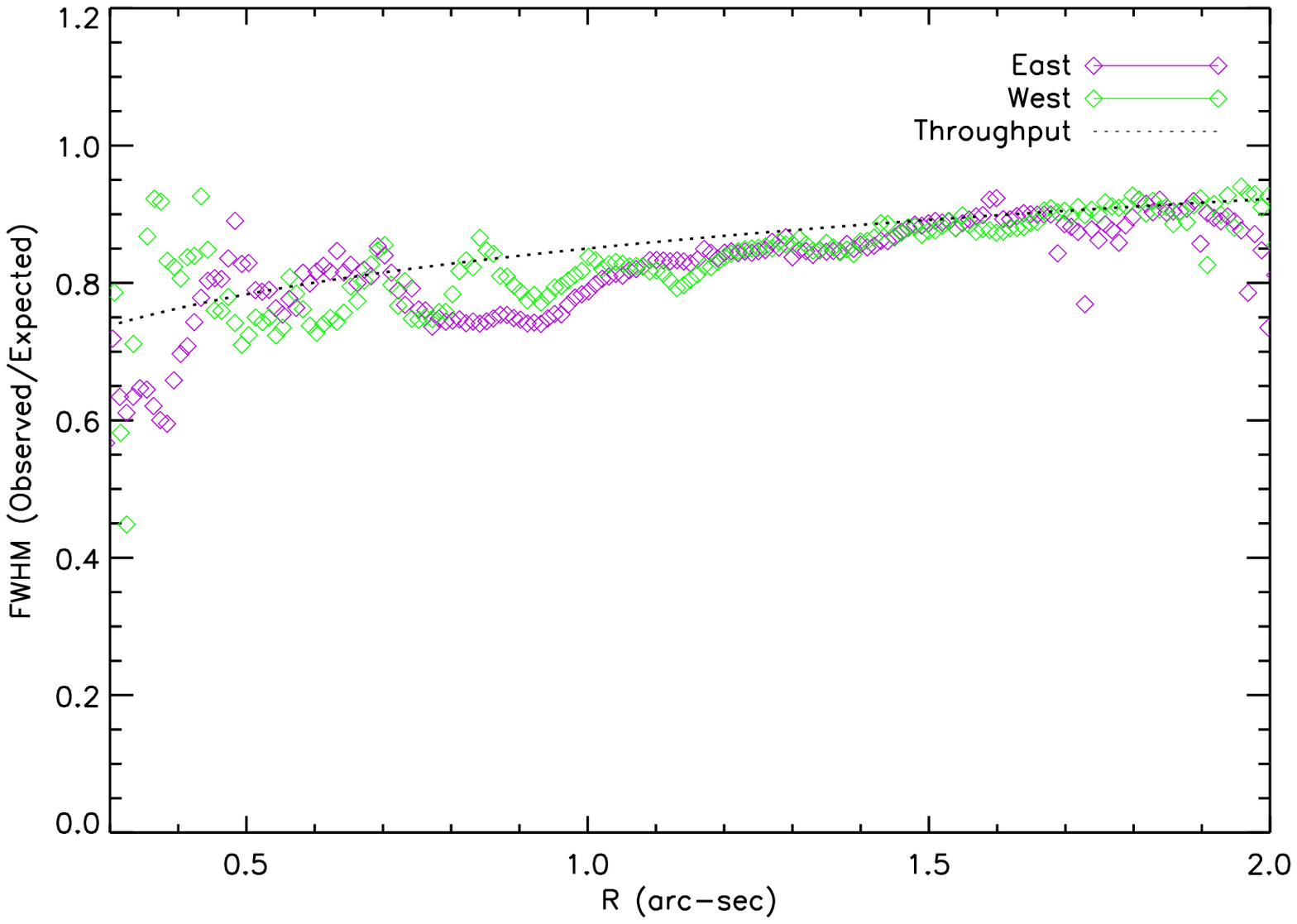}
\caption{(Top panels) Surface brightness (left) and disk FWHM (right) vs. angular separation for the 
fake disk.  (Bottom panels) Ratio of the ``observed" (after processing) and expected surface brightness (left) 
and disk FWHM (right) vs. angular separation.  The dotted lines identify power-law fits to correct our 
SB and disk FWHM measurements for biasing.}
\label{locidisksub_sbfwhm}
\end{figure}

\begin{figure}
\plotone{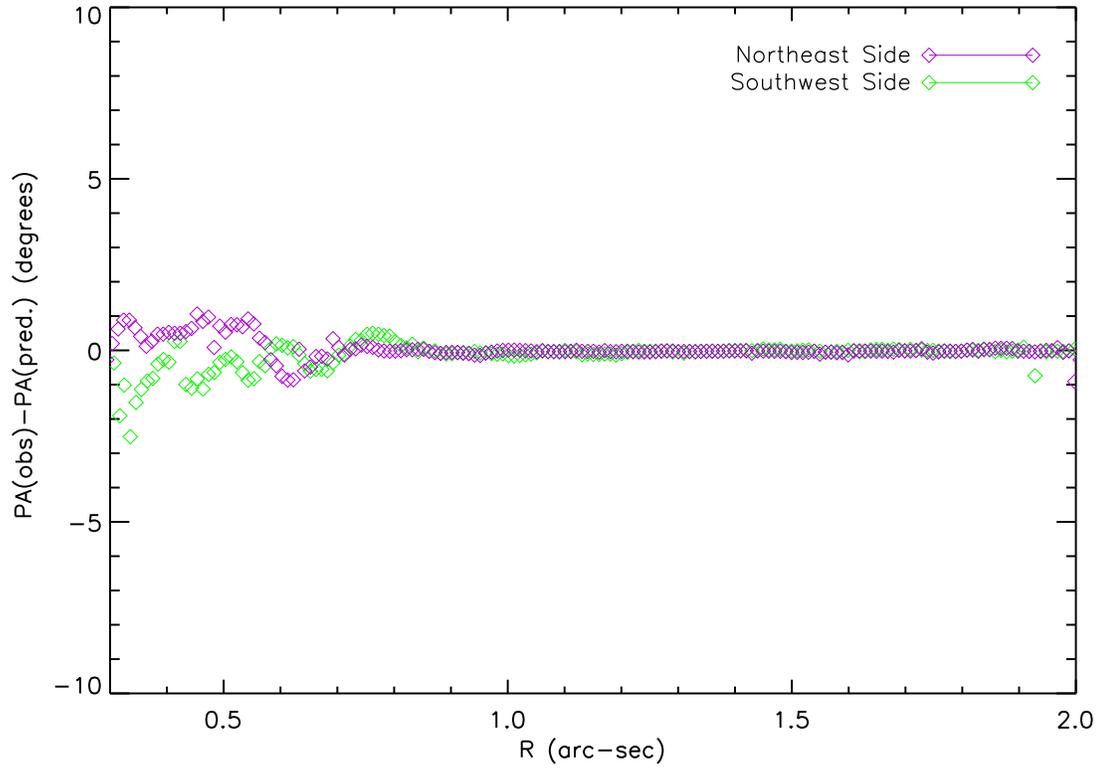}
\caption{Difference in position angle for the fake disk before and after processing.   Our processing does not bias the 
disk astrometry by more than 1--2$^{\circ}$ at any separation.}
\label{locidisksub_pa}
\end{figure}
\clearpage
\appendix
\section{Correcting for Disk Photometric/Astrometric Biases with LOCI}
The LOCI-based PSF subtraction approach can bias the photometry and astrometry of 
point sources like planets and extended structure like disks \citep[][]{Lafreniere2007,Thalmann2011}.  
For point source companions, imputing fake companions into registered images at a range of angular separations, 
processing these images, and then comparing the output to input fluxes and 
centroid positions of the fake sources corrects for these biases.  
Correcting biases for disks not viewed perfectly edge-on/pencil-thin is more 
difficult since we do not know, a priori, the disk's true FWHM, our ignorance of 
which also limits our ability to 'debias' other disk properties whose determination depends on the 
disk FWHM (i.e. surface brightness).  
Here we describe our method for mitigating LOCI-based photometric/astrometric 
biases, which largely follows that of \citet{Rodigas2012}.  

We first construct model disks each with midplane brightnesses (and thus signal-to-noise ratios) 
about twice that of the processed, real disk image and surface brightness profiles comparable to the real disk 
profiles beyond $r$ = 1\arcsec{}.  We consider two FWHM values of 0.2\arcsec{} and 0.25\arcsec{} 
at 1\arcsec{}, or $\sim$ 20--40\% larger than the values we get between $r$ = 0.3 and 1\arcsec{} prior to applying the 
corrections we derive in this section (Figure \ref{locidisksub_sbfwhm}, top panels).  
For both runs, we imput the fake disk into registered images, 
rerun our processing pipeline and then compare the output and input surface brightness profiles, disk thicknesses, and 
disk position angles.

As shown in Figure \ref{locidisksub_sbfwhm} (middle and bottom panels), our processing minimially biases the 
thinner model disk (FWHM  = 0.2\arcsec{}) exterior to $r$ = 1\arcsec{}, reducing its surface brightness and thickness 
by no more than 20\%.  Interior to $r$ = 1\arcsec{}, the surface brightness and thickness drop to no less than 
60--75\% of their original values.  Biasing for the FWHM = 0.25\arcsec{} disk (not shown) is only slightly more severe, resulting in an additional
$\sim$ 5\% (10\%) drop at $r$ $>$ 1\arcsec{} ($r$ $<$ 1\arcsec{}) in SB and FWHM.   For both fake disks, LOCI minimally biases the disk position angle measurements 
(Figure \ref{locidisksub_pa}).

The ``observed" FWHM for the model with the initially thinner disk (FWHM =0.2\arcsec{}) 
is $\sim$ 10\% smaller than the ``observed FWHM for the initially thicker disk (FWHM = 0.25\arcsec{}) and more similar to that 
which we find for the real HD 32297 disk prior to applying any bias corrections.  Thus, we derive throughput corrections for the 
disk surface brightness and FWHM for the thinner disk, fitting a simple, unweighted power law to data between 
$r$ = 0.3\arcsec{} and $r$ = 1.5\arcsec{}.  

The deviations in throughput for individual points vs. our throughput function are as large as $\sim$ 10--20\%.
However, our modeling errors for the disk SB and FWHM are 
larger in regions where biasing from LOCI is important and at $r$ $>$ 1.5--2\arcsec{} where the disk 
is intrinsically much fainter.  Therefore, 
we leave a more detailed, robust calibration of disk parameters from LOCI-processed images to a future work where 
a higher SNR disk detection at $r$ $<$ 1\arcsec{} will help improve our fitting precision.


\begin{thebibliography}{}
\bibitem[Augereau et al.(2001)]{Augereau2001}Augereau, J. C., 2001, A\&A, 370, 440
\bibitem[Backman et al.(1992)]{Backman1992}Backman, D., et al., 1992, \apj, 385, 670
\bibitem[Beuzit et al.(2008)]{Beuzit2008}Beuzit, J.-L., et al., 2008, SPIE, 7014, 41
\bibitem[Buenzli et al.(2010)]{Buenzli2010}Buenzli, E., et al., 2010, A\&A, 524, 1L
\bibitem[Currie et al.(2008)]{Currie2008}Currie, T., et al., 2008, \apj, 672, 558
\bibitem[Currie et al.(2009)]{Currie2009}Currie, T., Lada, C. J, et al., 2009, \apj, 698, 1
\bibitem[Currie et al.(2010)]{Currie2010}Currie, T., et al., \apj, 2010, 721, 177L
\bibitem[Currie et al.(2011a)]{Currie2011a}Currie, T., Burrows, A., et al., 2011a, \apj, 729, 128
\bibitem[Currie et al.(2011b)]{Currie2011b}Currie, T., Thalmann, C., et al., 2011b, \apj, 734, 36L
\bibitem[Currie et al.(2012)]{Currie2012}Currie, T., Fukagawa, M., et al., 2012, \apj\ submitted, arXiv:1206.0483
\bibitem[Dalle Ore et al. (2011)]{dalleore11}Dalle Ore, C.~M., et al., 2011, A\&A, 533, 98
\bibitem[Dawson et al.(2011)]{Dawson2011}Dawson, R., Murray-Clay, R., Fabrycky, D., 2011, \apj, 743, 17L
\bibitem[Debes et al.(2008)]{Debes2008}Debes, J. H., et al., 2008, \apj, 684, 41L
\bibitem[Debes et al.(2009)]{Debes2009}Debes, J. H., et al., 2009, \apj, 702, 318
\bibitem[Esposito et al.(2011)]{Esposito2011}Esposito, S., et al., 2011, SPIE, 8149, 1
\bibitem[Fitzgerald et al.(2007)]{Fitzgerald2007}Fitzgerald, M., et al., 2007, \apj, 670, 557
\bibitem[Fitzgerald et al.(2010)]{Fitzgerald2010}Fitzgerald, M., et al., 2010, Proceedings of the conference \textit{In the Spirit of Lyot 2010: Direct Detection of Exoplanets and Circumstellar Disks}, October 25 - 29, 2010, University of Paris Diderot, Paris, France. Ed. Anthony Boccaletti.
\bibitem[Golimowski et al.(2006)]{Golimowski2006}Golimowski, D., et al., 2006, \aj, 131, 3109
\bibitem[Heap et al.(2000)]{Heap2000}Heap, S., et al., 2000, \apj, 539, 435
\bibitem[Hines et al.(2007)]{Hines2007}Hines, D., et al., 2007, \apj, 671, 165L
\bibitem[Hong et al.(1985)]{hong85}Hong, S.~S., 1985, A\&A, 146, 67
\bibitem[Kalas et al.(2005)]{Kalas2005a} Kalas, P., Graham, J., and Clampin, M., 2005, Nature, 435, 1067 
\bibitem[Kalas(2005)]{Kalas2005b}Kalas, P., 2005b, \apj, 635, 169L
\bibitem[Kalas et al.(2007)]{Kalas2007}Kalas, P., et al. 2007, \apj, 661, 85L 
\bibitem[Kalas et al.(2008)]{Kalas2008}Kalas, P., et al., 2008, Science, 322, 1345
\bibitem[Kenyon and Bromley(2008)]{KenyonBromley2008}Kenyon, S. J., Bromley, B., 2008, \apjs, 179, 451
\bibitem[Kuchner and Holman(2003)]{Kuchner2003}Kuchner, M., Holman, M., 2003, \apj, 588, 1110
\bibitem[Kuchner and Stark(2010)]{KuchnerStark2010}Kuchner, M., Stark, C., 2010, \aj, 140, 1007
\bibitem[Lafreniere et al.(2007)]{Lafreniere2007}Lafreniere, D., et al., 2007, \apj, 660, 770
\bibitem[Lagrange et al.(2010)]{Lagrange2010}Lagrange, A.-M., et al., 2010, Science, 329, 57
\bibitem[Liou and Zook(1999)]{Liou1999}Liou, J. C., Zook, H., 1999, \aj, 118, 580
\bibitem[Macintosh et al.(2008)]{Macintosh2008}MacIntosh, B., et al., 2008, SPIE, 7015, 31
\bibitem[Martinache and Guyon(2009)]{Martinache2009}Martinache, F., Guyon, O., 2009, SPIE, 7440, 20
\bibitem[Maness et al.(2008)]{Maness2008}Maness, H., et al., 2008, \apj, 686, 25L
\bibitem[Marois et al.(2006)]{Marois2006}Marois, C., et al., 2006, \apj, 641, 556
\bibitem[Marois et al.(2008)]{Marois2008}Marois, C., et al., 2008, Science, 322, 1348
\bibitem[Marois et al.(2010)]{Marois2010}Marois, C., et al., 2010, Nature, 468, 1080
\bibitem[Mawet et al.(2009)]{Mawet2009}Mawet, D., et al., 2009, 702, 47L
\bibitem[Metchev et al.(2009)]{Metchev2009}Metchev, S., Marois, C., Hillenbrand, L., 2009, \apjs, 181, 62
\bibitem[Moerchen et al.(2007)]{Moerchen2007}Moerchen, M., et al., 2007, \apj, 666, 109L
\bibitem[Mulders et al.(2012)]{Mulders2012}Mulders, G., et al., 2012, A\&A\ submitted
\bibitem[Murakami et al.(2007)]{Murakami2007}Murakami, H., et al., 2007, PASJ, 59, 369
\bibitem[Plavchan et al.(2009)]{Plavchan2009}Plavchan, P., et al., 2009, \apjs, 698, 1068
\bibitem[Press et al.(1992)]{Press1992}Press, W., et al., 1992, \textit{Numerical Recipes in 
Fortran: The Art of Scientific Computing}, Cambridge: University Press, 1992, 2nd ed.
\bibitem[Quillen et al.(2006)]{Quillen2006}Quillen, A., 2006, \mnras, 372, 14L
\bibitem[Rhee et al.(2007)]{Rhee2007}Rhee, J., et al., 2007, \apj, 660, 1556
\bibitem[Rodigas et al.(2011)]{Rodigas2011}Rodigas, T. J., et al,. 2011, \apj, 732, 32L
\bibitem[Rodigas et al.(2012)]{Rodigas2012}Rodigas, T. J., et al., 2012, \apj\ in press
\bibitem[Schneider et al.(2005)]{Schneider2005}Schneider, G., et al., 2005, 629, 117L
\bibitem[Skemer et al.(2012)]{Skemer2012}Skemer, A., et al., 2012, \apj\ submitted
\bibitem[Smith and Terrile(1984)]{Smith1984}Smith, B., Terrile, R. J., 1984, Science, 226, 1421
\bibitem[Stark and Kuchner(2008)]{Stark2008}Stark, C., Kuchner, M., 2008, \apj, 686, 637
\bibitem[Su et al.(2009)]{Su2009}Su, K., et al., 2009, \apj, 705, 314
\bibitem[Thalmann et al.(2011)]{Thalmann2011}Thalmann, C., et al., 2011, \apj, 743, 6L
\bibitem[van Leeuwen(2007)]{vanLeeuwen2007}van Leeuwen, F, 2007, A\&A, 474, 653
\bibitem[Werner et al.(2004)]{Werner2004}Werner, M., et al., 2004, \apjs, 154, 1
\bibitem[Wright et al.(2010)]{Wright2010}Wright, E. L., et al., 2010, \aj, 140, 1868
\bibitem[Wyatt et al.(1999)]{Wyatt1999}Wyatt, M., et al., 1999, \apj, 527, 918
\bibitem[Wyatt et al.(2006)]{Wyatt2006}Wyatt, M., et al., 2006, \apj, 639, 1153
\bibitem[Wyatt(2008)]{Wyatt2008}Wyatt, M., 2008, \araa, 46, 339
\bibitem[Yelda et al.(2010)]{Yelda2010}Yelda, S., et al., 2010, \apj, 725, 331
\end{thebibliography}
\end{document}